\documentclass[12pt,preprint]{aastex}

\slugcomment{}

\shorttitle{PAHs in Elliptical Galaxies}
\shortauthors{Kaneda et al.}

\begin{document}


\title{Properties of polycyclic aromatic hydrocarbons in local elliptical galaxies revealed by the Infrared Spectrograph on Spitzer}


\author{H. Kaneda\altaffilmark{1}, T. Onaka\altaffilmark{2}, I. Sakon\altaffilmark{2}, T. Kitayama\altaffilmark{3}, Y. Okada\altaffilmark{1}, and T. Suzuki\altaffilmark{1}}

\altaffiltext{1}{Institute of Space and Astronautical Science, Japan Aerospace Exploration Agency, Sagamihara, Kanagawa 229-8510, Japan; kaneda@ir.isas.jaxa.jp, okada@ir.isas.jaxa.jp.}
\altaffiltext{2}{Department of Astronomy, Graduate School of Science, University of Tokyo, Bunkyo-ku, Tokyo 113-0003, Japan; onaka@astron.s.u-tokyo.ac.jp, isakon@astron.s.u-tokyo.ac.jp.}
\altaffiltext{3}{Department of Physics, Toho University, Funabashi, Chiba 274-8510, Japan; kitayama@ph.sci.toho-u.ac.jp.}




\begin{abstract}
We performed mid-infrared spectroscopic observations of 18 local dusty elliptical galaxies by using the Infrared Spectrograph (IRS) on board {\it Spitzer}. We have significantly detected polycyclic aromatic hydrocarbon (PAH) features from 14 out of the 18 galaxies, and thus found that the presence of PAHs is not rare but rather common in dusty elliptical galaxies. Most of these galaxies show an unusually weak 7.7 $\mu$m emission feature relative to 11.3 $\mu$m and 17 $\mu$m emission features. A large fraction of the galaxies also exhibit H$_2$ rotational line and ionic fine-structure line emissions, which have no significant correlation with the PAH emissions. The PAH features are well correlated with the continuum at 35 $\mu$m, whereas they are not correlated with the continuum at 6 $\mu$m. We conclude that the PAH emission of the elliptical galaxies is mostly of interstellar origin rather than of stellar origin, and that the unusual PAH interband strength ratios are likely to be due to a large fraction of neutral to ionized PAHs.     
\end{abstract}


\section{Introduction}
Elliptical galaxies provide dust grains with a unique environment, i.e. irradiation by old stellar radiation fields with little ultraviolet (UV) light and interstellar media mostly dominated by hot plasma. Their evolution and survival in hot plasma environments are interesting and important problems that have been addressed by many authors (e.g. Draine \& Salpeter 1979; Dwek \& Arendt 1992; Tielens et al. 1994; Tsai \& Mathews 1995; Yamada \& Kitayama 2005). The presence of a considerable amount of interstellar dust in elliptical galaxies is quite common (e.g. Knapp et al. 1989). Some elliptical galaxies contain surprisingly large dust masses considering their sputtering destruction in the hot plasma environments (Goudfrooij \& de Jong 1995; Temi et al. 2004). 

It has been found that even polycyclic aromatic hydrocarbons (PAHs) are present in several local elliptical galaxies with {\it Spitzer} (J. D. Bregman et al. 2006; J. N. Bregram et al. 2006; Bressan et al. 2006; Kaneda et al. 2005; 2007a) and {\it AKARI} (Kaneda et al. 2007b). Some of their spectra exhibit unusual PAH interband strength ratios: usually strong emission features at 6.2, 7.7, and 8.6 $\mu$m are weak in contrast to prominent features at 11.3 $\mu$m. The apparent weakness of the 7.7 $\mu$m feature is significantly affected by the continuum underlying shortward of 10 $\mu$m (J. D. Bregman et al. 2006); once corrected for the continuum, some of the ratios become normal while some of the galaxies still show a weak 7.7 $\mu$m band. The weak PAH 6--9 $\mu$m emission might reflect physical conditions of the interstellar medium (ISM) in evolved systems such as elliptical galaxies, i.e. a dominance of neutral PAHs over ionized ones due to very soft radiation fields from evolved stars. Neutral PAHs emit significantly less in the 6--8 $\mu$m emission (Allamandola et al. 1989; Joblin et al. 1994). Then what is the supplying source of the PAHs in elliptical galaxies? Although true nature and characteristics of interstellar PAHs are still in debate, PAH emission is generally regarded as excellent tracer of star-forming activity (e.g. Peeters et al. 2004). In addition, PAHs are expected to be easily destroyed through collision with high energy particles in hot plasma, although reliable estimates on their destruction efficiency are still unavailable. Hence the presence of the PAHs in the elliptical galaxies seems to be incompatible with their passive interstellar environments dominated by hot plasma. 

There are mainly three candidates for the origin of PAHs in elliptical galaxies: products from stellar mass loss, recent merger events, and cooling flows. {\it ISO} results show that evolved stars can produce unusually strong 11.3 $\mu$m PAH features (Hony et al. 2001; Cesarsky et al. 1998), which may be attributed to the above unusual PAH interband strength ratios. On the other hand, J. N. Bregman et al. (2006) reported that normal quiescent elliptical galaxies do not exhibit significant PAH emission features; the presence of PAHs might suggest star formation in the recent past that has been triggered by a merger event (Kaneda et al. 2007a). In connection with the cooling flow, Temi et al. (2007) recently suggested that the AGN-assisted feedback outflow from a central reservoir may play an important role in supplying dust into the interstellar space of elliptical galaxies; PAHs might come from residual dust fragments diminished by sputtering.  

Alternatively, the characteristics of the PAH emission may be regulated by nuclear activities of the galaxies. A significant fraction of local elliptical and spiral galaxies are known to harbor a low-luminosity AGN (LLAGN), including a low-ionization nuclear emission region (LINER) nucleus (e.g. Ho et al. 1997). Smith et al. (2007) have reported that the peculiar PAH features with an unusually low ratio of the 7.7 $\mu$m to the 11.3 $\mu$m emission strength arise from systems with relatively hard radiation fields of LLAGNs. Sturm et al. (2006) have shown that IR-faint LINER galaxies have very weak PAH features in the 5--10 $\mu$m range but strong 11.3 $\mu$m features. 

Spectroscopic observations of elliptical galaxies are comparatively few in the mid-infrared (IR), and thus the properties of PAHs in such evolved systems as elliptical galaxies are less understood. For further discussion, a larger sample was definitely required to obtain the correct and comprehensive view. 
Hence we performed mid-IR spectroscopic observations of 18 nearby dusty elliptical galaxies by using the Infrared Spectrograph (IRS; Houck et al. 2004) on board {\it Spitzer} (Werner et al. 2004) within a framework of the {\it Spitzer} 3rd Guest Observers (GO3) program (PI: H.K.). The objective of the program is to investigate environmental effects on PAHs and very small grains (VSGs) in elliptical galaxies by increasing a number of the sample of such elliptical galaxies as observed in our GO1 program, where we detected the unusually low PAH 7.7/11.3 ratios (Kaneda et al. 2005). The sample galaxies are among {\it IRAS} dusty elliptical galaxies as described in Goudfrooij \& de Jong (1995); dust masses derived from the {\it IRAS} flux densities exceed by more than one order of magnitude the threshold where dust is replenished by stellar mass loss at the rate given by Faber \& Gallagher (1976) and destroyed for the destruction time scale of $10^7$ yr. This time scale corresponds to a typical maximum value for the sputtering destruction in elliptical galaxies known to contain X-ray-emitting gas (e.g. Canizares et al. 1987). The properties of the observed elliptical galaxies are summarized in Table.1, which have similar properties in morphological type, distance, size, and optical luminosity. Most of them are also detected in the X-ray with {\it ROSAT} (O'Sullivan et al. 2001), and thus relatively X-ray-bright dusty elliptical galaxies. 

The IRS spectra are ideal to study the environmental effects on PAHs; its continuous spectral range is well matched to study the overall properties of the PAHs emission features at 6--19 $\mu$m. We have significantly detected PAH emission features from 14 out of the 18 galaxies, most of which tend to show the aforementioned unusually low PAH 7.7/11.3 ratios as well as relatively strong 17 $\mu$m emission features. The detection of such PAH features will impose strong restrictions on the physical state of the PAHs and thus their origin. This will provide key information on the issues about the evolution of elliptical galaxies as well. 

\section{Observations and Data Reduction}
The {\it Spitzer}/IRS observations of the 18 nearby elliptical galaxies were carried out in our GO3 program from 2006 Jul to 2007 Aug, except for the IRS/Short-Low (SL) observations of the 7 galaxies that had been performed in our GO1 program. The summary of the observation log is listed in Table 2. For each galaxy, we obtained 5--37 $\mu$m spectra using the IRS/SL and Long-Low (LL) modules in the standard staring mode. Each target was observed as two fixed cluster offsets, one of which is positioned on the center of a galaxy, while the other is dedicated for observing its nearby blank sky. The sample consists of relatively dusty elliptical galaxies, and yet, even those elliptical galaxies are so faint in the mid-IR that we have taken long exposure in staring rather than short exposure in spectral mapping for the same total amount of observation time.

Starting from the S15.3 or S16.1 pipeline products provided by the {\it Spitzer} Science Center (SSC), we extracted and calibrated the IRS spectra by using the SPICE (version 2.0.4) software package. We first operated subtraction in two dimensions of background from signal data for each sub-slit and nod position; a blank sky was observed with a spatial separation of about $2'-3'$ from the center of each galaxy. The subtraction of the background resulted in removing many rogue pixels. Any remaining rogue pixels were then removed as much as possible by using the IRSCLEAN tool. 

As shown in Table 1, the galaxies are significantly extended more than IRS point sources. Therefore, we followed the reduction strategy for extended sources recommended by the SSC. We performed a 14-pixel-fixed column extraction on the sources along the diffraction orders, which corresponds to the extraction aperture lengths of $25''$ for SL and $71''$ for LL. These cover most of the mid-IR brightness distribution along the slit length for each galaxy. To estimate aperture-loss correction factors, we performed a similar extraction on a standard calibration star, HR~7341, and compared spectra to those obtained from the standard column extraction for point sources. In the diffraction direction, we assumed constant brightness distributions of the galaxies, because the slit widths were so short as compared to the galaxy sizes, by which slit-loss correction factors were taken into account. The resulting flux correction factors have wavelength-dependent values of 0.6--0.8 for SL and 0.7--0.9 for LL. Further division of the corrected flux density by the extracted area given by the slit width and the extraction aperture yields spectra in surface brightness units, i.e. MJy sr$^{-1}$. 
 
For each of the SL and LL modules, the third ``bonus'' order has not been included; the first and second orders matched fairly well and required no further base-line adjustment. The spectral mismatches caused by the difference in extraction aperture and slit width between the SL and LL modules were removed by scaling SL to match LL at several overlapping pixels. The ends of each order where the noise increases significantly were manually clipped. Figure 1 shows the mid-IR spectra thus obtained with the IRS for the 18 elliptical galaxies. In order to facilitate comparison, we use a common vertical axis range of a factor of 30 for each spectrum. For IC~3370, the SL observation missed the center of the galaxy by $10''$, which caused significant reduction in the signal-to-noise (S/N) ratio of the SL spectrum. 
 
\section{Results}
As shown in Fig.1, the PAH emission features are detected from most of the sample elliptical galaxies, and thus it is found that the presence of PAHs is not rare but rather common in dusty elliptical galaxies. Most of these galaxies have a PAH emission spectrum with an apparently unusually low ratio of the 7.7 $\mu$m to the 11.3 $\mu$m emission strength, as found in Kaneda et al. (2005). As described below, however, there will be possible problems with the measurement of the 7.7 $\mu$m band strength due to the presence of an underlying continuum. The PAH 17 $\mu$m emission appears to be strong relative to the PAH 7.7 $\mu$m emission as well. A large fraction of the galaxies also exhibit H$_2$ rotational lines and ionic fine-structure lines. At shorter wavelengths, the continua are dominated by blackbody emission from low-mass asymptotic giant branch (AGB) stars with temperatures of 3000 K to 5000 K (Athey et al. 2002; Xilouris et al. 2004; Temi et al. 2005; 2008). Interstellar dust contributes to the continuum emission at longer wavelengths through stochastic heating of VSGs. Only the gas emission lines of H$_2$S(1) and S(3), [NeII] 12.8$\mu$m, [NeIII] 15.6$\mu$m, and [SiII] 34.8$\mu$m are bright enough to obtain their line intensities from a significant fraction of the sample galaxies. In order to perform a statistical study, therefore, we concentrate on these gas emission lines as well as the three major PAH emission features at 7.7$\mu$m, 11.3$\mu$m, and 17 $\mu$m. 

In order to fit the full 5--37 $\mu$m low-resolution spectrum, we have used the PAHFIT program developed by the {\it Spitzer} SINGS legacy team (Smith et al. 2007; Kennicutt et al. 2003), which is primarily designed for decomposing {\it Spitzer} IRS spectra of PAH emission sources. The fitting results are summarized in Table 3. The strengths of the three main PAH bands are obtained by adding emission sub-features at 7.4, 7.6, and 7.8 $\mu$m for the 7.7 $\mu$m complex, those at 11.2 and 11.3 $\mu$m for the 11.3 $\mu$m complex, and those at 16.4, 17.0, 17.4, and 17.9 $\mu$m for the 17 $\mu$m complex. The PAH interband strength ratios of 7.7/11.3 and 17/11.3 as well as the gas line ratios of [NeIII]/[NeII] and H$_2$S(3)/H$_2$S(1) are listed in Table 4.

Most of the spectra of the sample galaxies are well reproduced by the PAHFIT program. However, even by taking into account that spectral data errors are only statistical ones, reduced chi-squares for fitting are unacceptably high in some galaxies, where the spectrum contains silicate emission as silicate dust signature of AGB star winds (e.g. NGC~1407; Bressan et al. 2006), or a strong mid-IR continuum emitted by hot dust (NGC~1052). Such contributions are not considered in the PAHFIT program. Therefore, for the spectral ranges affected by those features, we locally fitted the spectrum by quadratic base-line fits; we used Gaussians to fit the line profile of atomic/H$_2$ lines and Lorentzians for discrete PAH features. As a result, we have confirmed that the detection of the PAH 11.3 $\mu$m feature is significant for all the galaxies except NGC~1407, NGC~1549, NGC~3904, and NGC~4696 with a S/N ratio larger than 5 (Table 3).

As pointed out by J. D. Bregman et al. (2006), a broad 7.7 $\mu$m emission feature may appear to be relatively weak due to the presence of a stellar silicate feature and continuum underlying shortward of 10 $\mu$m. In order to investigate the effect, we have obtained an approximate spectral template of the stellar component by averaging the spectra of the three quiescent elliptical galaxies in the sample, NGC~1407, NGC~1549, and NGC~3904, and then subtracted the template from each spectrum in Fig.1, normalized at the wavelength of $\sim$5.4 $\mu$m. The spectra thus obtained are shown in Fig.2 for the representative elliptical galaxies. The broad 7.7 $\mu$m feature that can hardly be recognized in Fig.1 emerges clearly in Fig.2, which is due to the removal of the continuum depression around 8 $\mu$m as seen in the spectra of NGC~1407, NGC~1549, and NGC~3904 (Fig.1). We have again used the PAHFIT program to obtain the strength of the PAH 7.7$\mu$m feature for the stellar-component-subtracted spectra, which will be denoted as PAH 7.7$^*$ hereafter. The resultant PAH 7.7$^*$ as well as the PAH 7.7$^*$/11.3 ratio are summarized in Table 5. The bottom panel in Fig.2 shows comparison between PAH 7.7$^*$ and PAH 7.7; in the high and intermediate ranges of the PAH 7.7$\mu$m feature strength, the difference between PAH 7.7 and 7.7$^*$ is only a factor of about 1.5--2, however PAH 7.7$^*$ can be 7--8 times stronger than PAH 7.7 in the low region. Hence we consider PAH 7.7$^*$ instead of PAH 7.7 in the following discussion.  

\section{Discussion}
\subsection{Correlation plots of PAH 11.3 with other spectral components}
Figure 3 shows the variations of the strength of the PAH 11.3 $\mu$m feature with respect to those of other spectral components, which are derived from the spectral fitting. The continuum intensities at 6 $\mu$m and 35 $\mu$m are monochromatic values obtained from the best-fit model continuum. For each plot, we use a common vertical axis range of 3 orders of magintude. Their variations among the galaxies are quite different from component to component, as seen in Fig.3. Particularly, the continuum at 6 $\mu$m shows relatively small variations from galaxy to galaxy. This is consistent with the similarity in the optical luminosity among the sample galaxies (Table 1), because the continua at 6 $\mu$m of elliptical galaxies are dominated by AGB stellar photospheric emission (Athey et al. 2002; Xilouris et al. 2004; Temi et al. 2005; 2008).

Here and hereafter, the box in the figures indicates that the galaxy have a LLAGN, categorized as LINER, Seyfert, or possibly LINER on the basis of the NASA IPAC extragalactic database (NED); in our sample, 9 galaxies out of 18 are such targets. The thick box in each figure corresponds to the plot for NGC~1052, where the continuum emission at long wavelengths is exceptionally strong (see Fig.1) and most probably carried by hot dust heated by its nuclear activities; NGC~1052 is a prototypical LINER galaxy (Sugai \& Malkan 2000). A linear-correlation coefficient, $R$, obtained by a straight line fit to the data points with S/N ratios greater than 3, is also attached to each plot, together with the number of the data points used in the calculation. We below adopt the criterion of $R>0.7$ as statistical significance of correlation, which implies confidence levels of about 95--99.5 \% for $8\sim14$ data points (Bevington \& Robinson 2003). For the plots related to the continuum emission at 35 $\mu$m, the contribution of NGC~1052 is excluded in calculating the correlation coefficient.  

As seen in Fig.3, the strength of the PAH 11.3 $\mu$m feature is significantly correlated with those of the PAH 7.7$^*$ $\mu$m and the continuum at 35 $\mu$m, while it is not correlated with the strengths of the PAH 17 $\mu$m feature, the [NeII]12.8 $\mu$m, and the H$_2$S(1) line as well as the continuum at 6 $\mu$m. It is not correlated with the [SiII]34.8 $\mu$m line strength, either ($R=0.096$). In particular, the PAH feature strength increases in proportion to the intensity of the interstellar VSG continuum at 35 $\mu$m. In contrast, there is no significant correlation of the PAH with the stellar continuum emission at 6 $\mu$m. If the PAHs were supplied by mass loss from evolved stars and subsequently destroyed in the hot plasma, the observed strengths of the PAH and the stellar emission would be closely linked to each other. Hence, the marked difference in the relation of the PAH with the 6 $\mu$m and the 35 $\mu$m continuum implies that the observed PAHs are mostly of interstellar origin rather than of stellar origin, supporting the result obtained by J. D. Bregman et al. (2006) from an elliptical galaxy NGC~4697. 
 
In addition to the PAH features, a series of H$_2$ rotational lines as well as ionic fine-structure lines such as [NeII] and [NeIII] are detected from a large fraction of the elliptical galaxies, implying the presence of warm molecular and ionized gases. Since there is no significant correlation between the strengths of the [NeII]/H$_2$ lines and the PAH feature (Fig.3), the PAH emission does not seem to come from the same regions that emit the [NeII] and H$_2$ lines. Thus the ISM of the elliptical galaxies seems to consist of components of a wide range of physical conditions. 

\subsection{PAH interband strength ratios}
Figure 4 shows the PAH 7.7$^*$/11.3 and PAH 17/11.3 ratios plotted against the strength of the PAH 11.3 $\mu$m feature. As for PAH 7.7$^*$/11.3, 12 out of 15 data points in Fig.4 have significantly ($>3\sigma$) small values as compared to the ratio of $\sim$4.2 that is the median of the galaxies with HII region or starburst-like optical spectra in the SINGS sample (Fig.14 in Smith et al. 2007). They also exhibit systematically lower values than 2.0--3.3 for Galactic diffuse emission (Sakon et al. 2004) and 1.8--3.0 for normal galaxies (Lu et al. 2003), both of which were estimated by spline-based continuum fitting. Because the full decomposition by PAHFIT produces about 2 times larger PAH 7.7/11.3 ratios than the conventional spline-based fitting (Smith et al. 2007), the differences are significant for most of the galaxies. On the other hand, the mean value ($\sim$0.6) of the PAH 17/11.3 ratios is close to the ratio of $\sim$0.5 averaged over all the SINGS sample galaxies, thus exhibiting usual interband strength ratios as a whole, although they have a large scatter. 

Figure 4 indicates that the variations of the PAH 7.7$^*$/11.3 and the PAH 17/11.3 ratio with the PAH 11.3 $\mu$m emission strength is not significant. The overall behavior of the PAH 17/11.3 ratio suggests that the size distribution of the PAHs is not biased toward large sizes without any systematic change with the PAH emission strength, because the PAH 17/11.3 ratio strongly depends on the size of PAHs (e.g. Draine \& Li 2007). On the other hand, the PAH 7.7/11.3 ratio strongly depends on the fraction of ionized to neutral PAHs (Allamandola et al. 1989; Draine \& Li 2007); the dependence of the 7.7/11.3 ratio on the PAH size is rather weak (Schutte et al. 1993). For example, changes of the ionizing state of PAHs by decrease in the PAH 7.7/11.3 ratio were observed in Galactic reflection nebulae with distance from the central star (Bregman \& Temi 2005), and from the arm to the inter-arm region in the spiral galaxy NGC~6946 (Sakon et al. 2007). Hence the unusual weakness of the PAH 7.7$\mu$m emission features in the elliptical galaxies is reasonably explained by a larger fraction of neutral to ionized PAHs due to very soft interstellar radiation field from evolved stars. 

As seen in Figs.3 and 4, there is no systematic difference in strengths of the PAH 7.7 and 11.3 $\mu$m features between the LLANG and the non-LLAGN elliptical galaxies, although the three faintest galaxies in both PAH and gas line emissions belong to non-LLAGNs. However, there seems to be a significant difference in the strength of the PAH 17$\mu$m feature. The galaxies with a LLAGN show stronger PAH 17$\mu$m features, suggesting that the AGN-assisted feedback flow may contribute to the supply of large PAHs into the interstellar space. As for NGC~5044 in our sample, Temi et al. (2007) found evidence of such feedback flow. They suggested the presence of PAHs in the flow on the basis of the IRAC $8-4.5$ $\mu$m difference image, which might come from residual dust fragments diminished by sputtering. In the IRS spectrum of NGC~5044 in Fig.1, there may be a broad PAH feature around 7.7 $\mu$m, although it is not significant due to a low S/N; for the stellar-continuum-subtracted spectrum in Fig.2, the 7.7 $\mu$m feature is significant with $\sim$1.5 times higher strength (Table 5). In addition, the H$_2$S(5) line at 6.9 $\mu$m is so strong that it contributes to the IRAC 8 $\mu$m (6.5--9$\mu$m) band by $\sim$2 \% based on the IRAC system respose curve in the {\it Spitzer} Observer's Manual with the observed spectrum.

\subsection{Relation of PAH ratios with gas line ratios}
To investigate the relation of the PAH emission features with the warm and the ionized gas, including the relation of their heating/excitation mechanisms, the variations of the PAH 7.7$^*$/11.3 and 17/11.3 ratios are plotted against the line ratios of [NeIII]/[NeII] and H$_2$S(3)/H$_2$S(1) in Fig.5. There are no significant correlation for every plot according to the criterion $R>0.7$.

The [NeIII]/[NeII] line ratio is a good indicator of hardness of radiation field with an ionization potential ratio of 41 eV/21.6 eV. As seen in Fig.5, the ratios are systematically higher for the LLAGN galaxies than those for the non-LLAGN galaxies. Therefore, the highly-ionized gas responsible for the [NeII] and [NeIII] lines is likely to be associated with the LLAGN (e.g. AGN tori).  Although our result is consistent with the plot in Fig.14 of Smith et al. (2007), in our observed range of [NeIII]/[NeII], we cannot obtain a significant correlation of the [NeIII]/[NeII] with either the PAH 17/11.3 or the PAH 7.7$^*$/11.3 ratio (Fig.5). Therefore the PAH emission features of the elliptical galaxies are not significantly affected by the hard radiation from the LLAGNs. In fact, the hard radiation field cannot permeate the diffuse ISM of the whole galaxy; otherwise, the 7.7 $\mu$m emission features from ionized PAHs would be dominant as usual.

The prototypical LINER galaxy, NGC~1052, shows the smallest [NeIII]/[NeII] line ratio among the LINER galaxies in the sample (Fig.5). On the basis of such a small [NeIII]/[NeII] line ratio observed with {\it ISO}, Sugai \& Malkan (2000) concluded that slow-shock excitation is a dominant source for NGC~1052. They obtained an upper limit of $<$0.3 for the [NeIII]/[NeII] line ratio. We have a better measurement, which shows a larger value (0.58$\pm$0.06) than their upper limit. Nevertheless both measurements are in agreement with each other at the 3$\sigma$ level. The other LINER galaxies show rather larger [NeIII]/[NeII] line ratios despite their low nuclear activities inferred from much weaker dust continuum emission at longer wavelengths (Fig.1). 

The presence of H$_2$ lines implies that there is the far-UV emission from early-type stars illuminating photodissociated regions (PDRs) or shocks providing collisional heating. PAHs are excited not only by far-UV photons but also by lower energy photons. As seen in Fig.5, the galaxies in the sample show the H$_2$S(3)/S(1) line ratios ranging from 0.6 to 3, which correspond to the H$_2$ excitation temperatures of $200-400$ K (Turner et al. 1977). 

The properties of warm molecular hydrogen were studied in detail from a series of pure rotational H$_2$ lines of the {\it Spitzer} SINGS galaxy sample (Roussel et al. 2007). They found that a large fraction of the targets classified as LINERs or Seyferts showed a clear departure from the rest of the sample in having warmer H$_2$ in the excited states, suggesting that there is a dominant contribution of the H$_2$ lines from large-scale shock heating by dynamical perturbation (e.g. the interaction of nuclear jet with the ISM). From this viewpoint, we do not see a clear difference between the LLAGN and the non-LLAGN elliptical galaxies in Fig.5, although there is a hint in Fig.3 that the LLAGN galaxies have stronger H$_2$S(1) emission with respect to the PAH 11.3$\mu$m emission.

\subsection{Dependence on X-ray luminosity}
In Fig.6, the equivalent widths of the PAH 7.7$^*$ $\mu$m, 11.3 $\mu$m, and 17 $\mu$m features as well as that of the [SiII] line are plotted as a function of the X-ray relative to the blue luminosity (Table 1). Here, since the X-ray luminosities of IC~3370 and NGC~7052 are not given in any paper, we estimate their upper limits on the basis of the $\log L_{\rm B}$ vs. $\log L_{\rm X}$ scatter plot in O'Sullivan et al. (2001). 

As seen in Fig.6, there is no significant correlation between the PAH strengths and the X-ray luminosity. The destruction of PAHs by X-ray-emitting hot plasma would make anti-correlations in these plots if the majority of the observed PAHs were constantly replenished into interstellar space through mass loss of AGB stars and balanced against destruction in very short time scales. Figure 6 shows no clear signature of interactions between the PAHs and hot plasma, which may be due to the fact that the observed PAHs were predominantly supplied by rather accidental events in the recent past encounter with a gas-rich galaxy or feed-back outflow from a center, and thus their current amounts are mostly determined by the time when such an event took place.

Incidentally, there is no significant correlation of the X-ray luminosity with the H$_2$S(1) line ($R=0.599$), the [NeII] line ($R=0.516$), and the ratio of the continuum at 35 $\mu$m to that at 6 $\mu$m ($R=0.273$), either. However, we have found a significant correlation between the X-ray luminosity and the [SiII] 34.8 $\mu$m line. Si$^{+}$ has a low ionization potential of 8.2 eV, and the [SiII] line is among important cooling lines of PDRs. From a sample of the {\it Spitzer} SINGS galaxies, the observed [SiII] power is too large to be explained by a PDR model, which implies either that the majority of the [SiII] emission comes from HII regions in high-metallicity nuclear regions or that a relatively small fraction of Si is incorporated in dust grains (Roussel et al. 2007). The [SiII] line can also be a significant coolant of X-ray-dominated regions associated with AGNs (Dale et al. 2006). However, the galaxies with LLAGNs in our sample do not show substantially larger [SiII] line strengths than those without LLAGNs. In addition, the X-ray luminosity here is obtained from the {\it ROSAT} catalog where spectra are fitted by a hot plasma model with temperature fixed at 1 keV (O'Sullivan et al. 2001), and therefore highly-absorbed hard X-rays from LLAGNs, if any, cannot account for such soft X-ray luminosity. Hence, the positive correlation of the [SiII] line with the X-ray luminosity may suggest relatively abundant gas-phase Si in interstellar hot plasma through sputtering destruction of the dust grains where Si is depleted. The difference from the case of the PAHs in which there is no apparent correlation with the X-ray might even indicate that the dominant supplying sources of silicate and carbonaceous grains are not the same. 

\section{Conclusions}
We performed mid-IR spectroscopic observations of 18 local dusty elliptical galaxies by using the {\it Spitzer}/IRS. We have significantly detected the PAH 11.3 $\mu$m emission feature from 14 out of the 18 galaxies, and thus found that the presence of PAHs is not rare but rather common in dusty elliptical galaxies. We have confirmed that the strength of the PAH 7.7 $\mu$m feature that appears to be extremely faint for some galaxies can be considerably underestimated unless the presence of a stellar silicate feature and continuum underlying shortward of 10 $\mu$m is taken into account. Even considering the effect of the stellar component, most of these galaxies show an unusually weak 7.7 $\mu$m emission feature relative to 11.3 $\mu$m and 17 $\mu$m emission features. A large fraction of the galaxies also exhibit H$_2$ rotational line and ionic fine-structure line emissions, which have no significant correlation with the PAH emissions. The PAH features are well correlated with the continuum at 35 $\mu$m, whereas they are not correlated with the continuum at 6 $\mu$m. We conclude that the PAH emission from the elliptical galaxies is mostly of interstellar origin rather than of stellar origin, and that the unusual PAH interband strength ratios are likely to be due to a large fraction of neutral to ionized PAHs. 

In this paper, we have taken a statistical approach to investigate the environmental effects on the PAHs in the elliptical galaxies, not concentrating on particular galaxies. We are currently conducting observations of the same sample elliptical galaxies in near-IR spectroscopy as well as 10-band photometry with {\it AKARI} (Murakami et al. 2007). We will report follow-on detailed studies of some interesting sample galaxies in seperate papers by combining {\it Spitzer} and {\it AKARI} results. 

\acknowledgments

This work is based on observations made with the {\it Spitzer} Space Telescope, which is operated by the Jet Propulsion Laboratory, California Institute of Technology under a contract with NASA. 
We would express our gratitude to the IRS team, the SSC, and the {\it Spitzer} SINGS legacy team for their dedicated work in generating the BCD and developing the PAHFIT program as well as the SPICE and IRSCLEAN software package. We would also express many thanks to the referee for giving us useful comments, particularly those about the underlying stellar continuum issue in deriving the strength of the PAH 7.7 $\mu$m feature. This work is financially supported by a Grant-in-Aid from the Ministry of Education, Culture, Sports, Science and Technology in Japan (No. 19740114).

\clearpage

\begin{deluxetable}{lrrrrrrr}
\tabletypesize{\scriptsize}
\tablecaption{Properties of the observed elliptical galaxies}
\tablewidth{0pt}
\tablehead{
\colhead{Name} & \colhead{Type} & \colhead{$D$} & \colhead{Size $D_{25}$} &\colhead{Log$L_{\rm B}$\tablenotemark{a}} & \colhead{Log$L_{\rm IR}$\tablenotemark{b}} & \colhead{Log$L_{\rm X}$\tablenotemark{c}}\\
 & & \colhead{(Mpc)} & \colhead{(arcmin)} &\colhead{(L$_{\odot}$)} & \colhead{(L$_{\odot}$)} & \colhead{(ergs s$^{-1}$)} 
}
\startdata
NGC~708 & E2  & 55.2 & 3.0$\times$2.5 & 10.74 & 10.15 & 43.03 \\
NGC~1052 & E4  & 17.7 & 3.0$\times$2.1 & 10.12 & 9.81 & 40.31 \\
NGC~1395 & E2  & 20.5 & 5.9$\times$4.5 & 10.36 & 8.45 & 40.89 \\
NGC~1407 & E0  & 20.6 & 3.5$\times$2.0 & 10.58 & 8.64 & 41.00 \\
NGC~1549 & E1  & 14.5 & 4.9$\times$4.1 & 10.28 & $<$7.72 & 39.92 \\
NGC~2974 & E4  & 28.3 & 3.5$\times$2.0 & 10.29 & 9.27 & 40.58 \\
NGC~3557 & E3  & 32.2 & 4.1$\times$3.0 & 10.76 & 9.34 & 40.58 \\
NGC~3894 & E4  & 46.4 & 2.8$\times$1.7 & 10.47 & 9.36 & 41.19 \\
NGC~3904 & E2  & 17.9 & 2.7$\times$1.9 & 10.06 & 8.76 & 40.74 \\
NGC~3962 & E1  & 21.7 & 2.6$\times$2.2 & 10.13 & 8.54 & $<$40.22 \\
IC~3370  & E2  & 39.1 & 2.9$\times$2.3 & 10.69 & 9.85 &$\dots$ \\
NGC~4589 & E2  & 24.6 & 3.2$\times$2.6 & 10.33 & 8.96 & 40.36 \\
NGC~4696 & E1p  & 37.0 & 4.5$\times$3.2 & 10.83 & 9.36 & 43.23 \\
NGC~5018 & E3  & 30.2 & 3.3$\times$2.5 & 10.57 & 9.75 & $<$40.53 \\
NGC~5044 & E0  & 30.2 & 3.0$\times$3.0 & 10.70 & 8.84 & 42.74 \\
NGC~5090 & E2  & 42.2 & 2.9$\times$2.4 & 10.41 & 9.67 & 41.49 \\
NGC~7052 & E4  & 58.1 & 3.3$\times$2.5 & 10.69 & 10.03 &$\dots$ \\
IC~1459 & E3  & 18.9 & 5.2$\times$3.8 & 10.37 & 9.17 & 40.71 \\
\enddata
\tablenotetext{a}{Total blue luminosity of the galaxies, derived using a solar absolute magnitude M$_{\rm bol,\odot} = +4.75$.}
\tablenotetext{b}{IR (1--500 $\mu$m) luminosity (Goudfrooij \& de Jong 1995 and references therein).} 
\tablenotetext{c}{X-ray luminosity from the $ROSAT$ catalog of O'Sullivan et al. (2001).}
\end{deluxetable}

\clearpage

\begin{deluxetable}{lccclcc}
\tabletypesize{\scriptsize}
\tablecaption{Observation log}
\tablewidth{0pt}
\tablehead{
\colhead{Name} & \colhead{R.A. (J2000)}&\colhead{Decl. (J2000)}&\colhead{Date} &\colhead{Observing mode} &\colhead{Exposure}&\colhead{Cycles}  
}
\startdata
NGC~708& $1^{\rm h}52^{\rm m}46.48^{\rm s}$&$36^{\rm d}09^{\rm m}06.6^{\rm s}$& 2006 Sep 18& IRS SL, staring&60 sec&4\\
        &&& 2006 Sep 18& IRS LL, staring&30 sec&6\\
NGC~1052& $2^{\rm h}41^{\rm m}04.80^{\rm s}$&$-8^{\rm d}15^{\rm m}20.7^{\rm s}$& 2006 Sep 11& IRS SL, staring&60 sec&4\\
        &&& 2006 Sep 11& IRS LL, staring&30 sec&4\\
NGC~1395& $3^{\rm h}38^{\rm m}29.79^{\rm s}$&$-23^{\rm d}01^{\rm m}39.7^{\rm s}$& 2005 Aug 08& IRS SL, staring&60 sec&4\\
        &&& 2007 Aug 31& IRS LL, staring&30 sec&4\\
NGC~1407 & $3^{\rm h}40^{\rm m}11.90^{\rm s}$&$-18^{\rm d}34^{\rm m}49.4^{\rm s}$& 2005 Aug 17& IRS SL, staring&60 sec&4\\
        &&& 2007 Aug 31& IRS LL, staring&30 sec&4\\
NGC~1549 & $4^{\rm h}15^{\rm m}45.13^{\rm s}$&$-55^{\rm d}35^{\rm m}32.1^{\rm s}$& 2006 Jul 25& IRS SL, staring&60 sec&8\\
        &&& 2006 Jul 25& IRS LL, staring&30 sec&8\\
NGC~2974 & $9^{\rm h}44^{\rm m}33.28^{\rm s}$&$-3^{\rm d}41^{\rm m}56.9^{\rm s}$& 2005 May 22& IRS SL, staring&60 sec&4\\
        &&& 2007 Jun 09& IRS LL, staring&30 sec&6\\
NGC~3557 & $11^{\rm h}09^{\rm m}57.65^{\rm s}$&$-37^{\rm d}32^{\rm m}21.0^{\rm s}$& 2007 Feb 12& IRS SL, staring&60 sec&6\\
        &&& 2007 Feb 12& IRS LL, staring&30 sec&6\\
NGC~3894 & $11^{\rm h}48^{\rm m}50.36^{\rm s}$&$+59^{\rm d}24^{\rm m}56.4^{\rm s}$& 2007 Feb 12& IRS SL, staring&60 sec&8\\
        &&& 2007 Feb 12& IRS LL, staring&30 sec&8\\
NGC~3904 & $11^{\rm h}49^{\rm m}13.22^{\rm s}$&$-29^{\rm d}16^{\rm m}36.3^{\rm s}$& 2007 Feb 12& IRS SL, staring&60 sec&6\\
        &&& 2007 Feb 12& IRS LL, staring&30 sec&8\\
NGC~3962 & $11^{\rm h}54^{\rm m}40.10^{\rm s}$&$-13^{\rm d}58^{\rm m}30.1^{\rm s}$& 2005 Jan 03& IRS SL, staring&60 sec&4\\
        &&& 2007 Jun 26& IRS LL, staring&30 sec&6\\
IC~3370  & $12^{\rm h}27^{\rm m}38.00^{\rm s}$&$-39^{\rm d}20^{\rm m}16.8^{\rm s}$& 2005 Feb 10& IRS SL, staring&60 sec&4\\
        &&& 2007 Jul 30& IRS LL, staring&30 sec&4\\
NGC~4589 & $12^{\rm h}37^{\rm m}25.03^{\rm s}$&$+74^{\rm d}11^{\rm m}30.8^{\rm s}$& 2004 Oct 20& IRS SL, staring&60 sec&4\\
        &&& 2006 Nov 16& IRS LL, staring&30 sec&6\\
NGC~4696 & $12^{\rm h}48^{\rm m}49.28^{\rm s}$&$-41^{\rm d}18^{\rm m}40.0^{\rm s}$& 2005 Feb 10& IRS SL, staring&60 sec&4\\
        &&& 2006 Aug 03& IRS LL, staring&30 sec&6\\
NGC~5018 & $13^{\rm h}13^{\rm m}01.00^{\rm s}$&$-19^{\rm d}31^{\rm m}05.1^{\rm s}$& 2006 Jul 25& IRS SL, staring&60 sec&8\\
        &&& 2006 Jul 25& IRS LL, staring&30 sec&6\\
NGC~5044 & $13^{\rm h}15^{\rm m}23.97^{\rm s}$&$-16^{\rm d}23^{\rm m}07.9^{\rm s}$& 2006 Jul 25& IRS SL, staring&60 sec&6\\
        &&& 2006 Jul 25& IRS LL, staring&30 sec&8\\
NGC~5090 & $13^{\rm h}21^{\rm m}12.81^{\rm s}$&$-43^{\rm d}42^{\rm m}16.4^{\rm s}$& 2006 Jul 29& IRS SL, staring&60 sec&8\\
        &&& 2006 Jul 29& IRS LL, staring&30 sec&6\\
NGC~7052 & $21^{\rm h}18^{\rm m}33.05^{\rm s}$&$+26^{\rm d}26^{\rm m}48.9^{\rm s}$& 2006 Dec 24& IRS SL, staring&60 sec&8\\
        &&& 2006 Dec 24& IRS LL, staring&30 sec&6\\
IC~1459 & $22^{\rm h}57^{\rm m}10.61^{\rm s}$&$-36^{\rm d}27^{\rm m}44.0^{\rm s}$& 2006 Nov 15& IRS SL, staring&60 sec&6\\
        &&& 2006 Nov 15& IRS LL, staring&30 sec&4\\
\enddata
\end{deluxetable}


\clearpage

\begin{deluxetable}{lrrrrrrrr}
\tabletypesize{\tiny}
\tablecaption{Intensities of gas emission lines and PAH features\tablenotemark{a}}
\tablewidth{0pt}
\tablehead{
\colhead{Name} &\colhead{[NeII]}&\colhead{[NeIII]}&\colhead{[SiII]}&\colhead{H$_2$S(1)}&\colhead{H$_2$S(3)}&\colhead{PAH 7.7}&\colhead{PAH 11.3}&\colhead{PAH 17} \\
\colhead{} &\multicolumn{8}{c}{($10^{-10}$ W m$^{-2}$ sr$^{-1}$)}\\
\colhead{} & EW\tablenotemark{b} ($\mu$m) & EW ($\mu$m) & EW ($\mu$m) & EW ($\mu$m) & EW ($\mu$m) & EW ($\mu$m) & EW ($\mu$m) & EW ($\mu$m)\\ 
}
\startdata
NGC~708&11.4$\pm$1.1&8.6$\pm$0.9&9.9$\pm$1.8&14.5$\pm$1.2&31.6$\pm$1.5&81$\pm$23&58$\pm$5&15$\pm$6\\
$\chi_{\nu}=1.34$&0.27&0.32&3.0&0.66&0.35&0.32&0.63&0.5\\
NGC~1052&106.0$\pm$4.6&61.5$\pm$5.7&116.0$\pm$2.7&19.8$\pm$2.1&25.4$\pm$2.7&77$\pm$12&75$\pm$5&110$\pm$20\\
$\chi_{\nu}=38.2$&0.067&0.043&0.277&0.015&0.014&0.025&0.030&0.06\\
NGC~1395&2.1$\pm$1.0&$<$3&$<$5&3.8$\pm$1.7&4.7$\pm$1.6&29$\pm$13&50$\pm$6&12$\pm$10\\
$\chi_{\nu}=1.31$&0.010&$<$0.02&$<$0.4&0.04&0.010&0.020&0.11&0.09\\
NGC~1407&$<$2&$<$3&$<$5&$<$3&$<$4&$<$10&$<$6&$<$5\\
$\chi_{\nu}=2.37$&$<$0.008&$<$0.02&$<$0.6&$<$0.02&$<$0.007&$<$0.005&$<$0.01&$<$0.03\\
NGC~1549&0.6$\pm$0.5&4.0$\pm$0.7&1.8$\pm$1.3&1.1$\pm$0.9&1.1$\pm$0.8&$<$10&$<$2&$<$6\\
$\chi_{\nu}=8.98$&0.004&0.029&0.2&0.011&0.02&$<$0.003&$<$0.004&$<$0.04\\
NGC~2974&24.8$\pm$0.9&17.2$\pm$1.0&33.6$\pm$2.1&14.5$\pm$1.7&24.5$\pm$1.2&160$\pm$46&207$\pm$5&130$\pm$9\\
$\chi_{\nu}=1.87$&0.099&0.105&0.79&0.11&0.053&0.13&0.42&0.65\\
NGC~3557&8.8$\pm$1.5&4.3$\pm$0.8&9.7$\pm$3.2&$<$2&4.0$\pm$1.0&41$\pm$7&73$\pm$4&19$\pm$8\\
$\chi_{\nu}=1.98$&0.11&0.036&0.5&$<$0.02&0.009&0.030&0.169&0.15\\
NGC~3894&13.7$\pm$0.6&14.2$\pm$0.8&15.5$\pm$1.6&10.8$\pm$1.0&9.7$\pm$1.0&224$\pm$38&80$\pm$4&30$\pm$7\\
$\chi_{\nu}=1.16$&0.087&0.130&0.36&0.12&0.046&0.34&0.30&0.22\\
NGC~3904&$<$2&$<$2&$<$10&$<$2&$<$2&$<$10&$<$3&$<$6\\  
$\chi_{\nu}=3.24$&$<$0.01&$<$0.01&$<$2&$<$0.04&$<$0.006&$<$0.006&$<$0.009&$<$0.08\\
NGC~3962&8.8$\pm$0.8&10.8$\pm$0.9&25.7$\pm$2.1&7.0$\pm$1.4&13.8$\pm$1.1&36$\pm$8&67$\pm$4&35$\pm$10\\
$\chi_{\nu}=1.20$&0.074&0.16&1.02&0.14&0.051&0.043&0.26&0.5\\
IC~3370&12.7$\pm$5.3&7.8$\pm$1.6&18.9$\pm$2.8&39.7$\pm$1.9&26.1$\pm$7.5&$<$150&121$\pm$23&160$\pm$15\\
$\chi_{\nu}=1.09$&0.06&0.06&0.33&0.35&0.05&$<$0.10&0.25&0.96\\
NGC~4589&7.4$\pm$1.0&6.4$\pm$0.9&14.4$\pm$2.0&13.5$\pm$1.3&8.2$\pm$3.3&$<$20&59$\pm$5&52$\pm$9\\
$\chi_{\nu}=1.05$&0.056&0.09&1.3&0.25&0.03&$<$0.02&0.21&0.6\\
NGC~4696&15.0$\pm$1.3&14.2$\pm$1.6&23.3$\pm$2.2&9.9$\pm$1.2&13.4$\pm$1.7&$<$10&16$\pm$6&14$\pm$6\\
$\chi_{\nu}=0.98$&0.15&0.25&5.6&0.23&0.064&$<$0.01&0.08&0.2\\
NGC~5018&20.4$\pm$0.7&14.9$\pm$1.0&11.5$\pm$6.2&9.4$\pm$1.5&26.0$\pm$5.0&601$\pm$33&380$\pm$4&129$\pm$12\\
$\chi_{\nu}=3.91$&0.073&0.085&0.06&0.06&0.043&0.35&0.656&0.59\\
NGC~5044&23.7$\pm$0.8&10.1$\pm$0.8&29.1$\pm$1.6&15.7$\pm$1.1&16.4$\pm$6.4&80$\pm$44&65$\pm$5&24$\pm$10\\
$\chi_{\nu}=1.12$&0.224&0.16&3.3&0.31&0.07&0.12&0.30&0.3\\
NGC~5090&18.8$\pm$0.6&11.3$\pm$1.7&4.2$\pm$1.5&$<$3&$<$2&$<$10&26$\pm$3&2$\pm$1\\ 
$\chi_{\nu}=2.01$&0.134&0.13&0.24&$<$0.03&$<$0.008&$<$0.01&0.10&0.02\\ 
NGC~7052&27.8$\pm$0.6&6.1$\pm$1.0&27.7$\pm$2.5&2.8$\pm$1.6&4.9$\pm$1.1&888$\pm$34&207$\pm$4&83$\pm$11\\
$\chi_{\nu}=2.11$&0.224&0.050&0.29&0.04&0.033&1.66&1.03&0.72\\
IC~1459&53.6$\pm$0.8&36.7$\pm$1.2&59.9$\pm$2.2&$<$2&4.0$\pm$1.0&10$\pm$9&54$\pm$5&79$\pm$10\\
$\chi_{\nu}=4.29$&0.139&0.157&0.59&$<$0.009&0.022&0.005&0.114&0.29\\ 
\enddata
\tablenotetext{a}{The uncertainies quoted for the line fluxes are the $1\sigma$ errors from the spectral fitting and do not include the calibration uncertainties. 2 $\sigma$ upper limits are given for the lines where the flux uncertainty exceeds the flux. }
\tablenotetext{b}{EW = equivalent width (observed).} 
\end{deluxetable}

\clearpage

\begin{deluxetable}{lrrrr}
\tabletypesize{\scriptsize}
\tablecaption{PAH interband strength ratios of 7.7/11.3 and 17/11.3 as well as the gas line ratios of [NeIII]/[NeII] and H$_2$S(3)/H$_2$S(1).}
\tablewidth{0pt}
\tablehead{
\colhead{Name} &\colhead{PAH 7.7/11.3}&\colhead{PAH 17/11.3}&\colhead{[NeIII]/[NeII]}&\colhead{H$_2$S(3)/H$_2$S(1)} \\
}
\startdata
NGC~708&1.4$\pm$0.4&0.26$\pm$0.11&0.75$\pm$0.11&2.2$\pm$0.2\\
NGC~1052&1.0$\pm$0.2&1.5$\pm$0.3&0.58$\pm$0.06&1.3$\pm$0.2\\
NGC~1395&0.58$\pm$0.27&0.24$\pm$0.20&$<$1.4&1.2$\pm$0.7\\
NGC~1407&$\dots$&$\dots$&$\dots$&$\dots$\\
NGC~1549&$\dots$&$\dots$&$\dots$&$\dots$\\
NGC~2974&0.77$\pm$0.22&0.63$\pm$0.05&0.69$\pm$0.05&1.7$\pm$0.2\\
NGC~3557&0.56$\pm$0.10&0.26$\pm$0.11&0.49$\pm$0.12&$>$2\\
NGC~3894&2.8$\pm$0.5&0.38$\pm$0.09&1.04$\pm$0.07&0.90$\pm$0.12\\
NGC~3904&$\dots$&$\dots$&$\dots$&$\dots$\\
NGC~3962&0.54$\pm$0.12&0.52$\pm$0.15&1.2$\pm$0.2&2.0$\pm$0.4\\
IC~3370&$<$1&1.3$\pm$0.3&0.61$\pm$0.29&0.66$\pm$0.19\\
NGC~4589&$<$0.3&0.88$\pm$0.17&0.86$\pm$0.17&0.61$\pm$0.25\\
NGC~4696&$<$0.6&0.88$\pm$0.50&0.95$\pm$0.13&1.4$\pm$0.2\\
NGC~5018&1.58$\pm$0.09&0.34$\pm$0.03&0.73$\pm$0.06&2.8$\pm$0.7\\
NGC~5044&1.2$\pm$0.7&0.37$\pm$0.16&0.43$\pm$0.04&1.0$\pm$0.4\\
NGC~5090&$<$0.4&0.08$\pm$0.04&0.60$\pm$0.09&$\dots$\\ 
NGC~7052&4.3$\pm$0.2&0.40$\pm$0.05&0.22$\pm$0.04&1.8$\pm$1.1\\
IC~1459&0.19$\pm$0.17&1.5$\pm$0.2&0.68$\pm$0.02&$>$2\\
\enddata
\end{deluxetable}

\clearpage

\begin{deluxetable}{lrr}
\tabletypesize{\scriptsize}
\tablecaption{Intensity of the PAH 7.7$^*$ feature and the PAH 7.7$^*$/11.3 ratio\tablenotemark{a}.}
\tablewidth{0pt}
\tablehead{
\colhead{Name} &\colhead{PAH 7.7$^*$}&\colhead{PAH 7.7$^*$/11.3} \\
}
\startdata
NGC~708&93$\pm$9&1.6$\pm$0.2\\
NGC~1052&90$\pm$20&1.2$\pm$0.3\\
NGC~1395&78$\pm$31&1.6$\pm$0.6\\
NGC~1407&$\dots$&$\dots$\\
NGC~1549&$\dots$&$\dots$\\
NGC~2974&316$\pm$27&1.5$\pm$0.1\\
NGC~3557&101$\pm$9&1.4$\pm$0.1\\
NGC~3894&290$\pm$14&3.6$\pm$0.3\\
NGC~3904&$\dots$&$\dots$\\
NGC~3962&111$\pm$22&1.7$\pm$0.3\\
IC~3370&168$\pm$58&1.4$\pm$0.6\\
NGC~4589&71$\pm$30&1.2$\pm$0.5\\
NGC~4696&$<$10&$<$0.6\\
NGC~5018&900$\pm$240&2.4$\pm$0.6\\
NGC~5044&117$\pm$19&1.8$\pm$0.3\\
NGC~5090&76$\pm$15&2.9$\pm$0.7\\ 
NGC~7052&959$\pm$35&4.6$\pm$0.2\\
IC~1459&83$\pm$12&1.5$\pm$0.3\\
\enddata
\tablenotetext{a}{The PAH 7.7$^*$ denotes the strength of the PAH 7.7$\mu$m feature obtained by fitting the stellar-component-subtracted spectra (see text).} 
\end{deluxetable}

\begin{figure}
\epsscale{.37}
\plotone{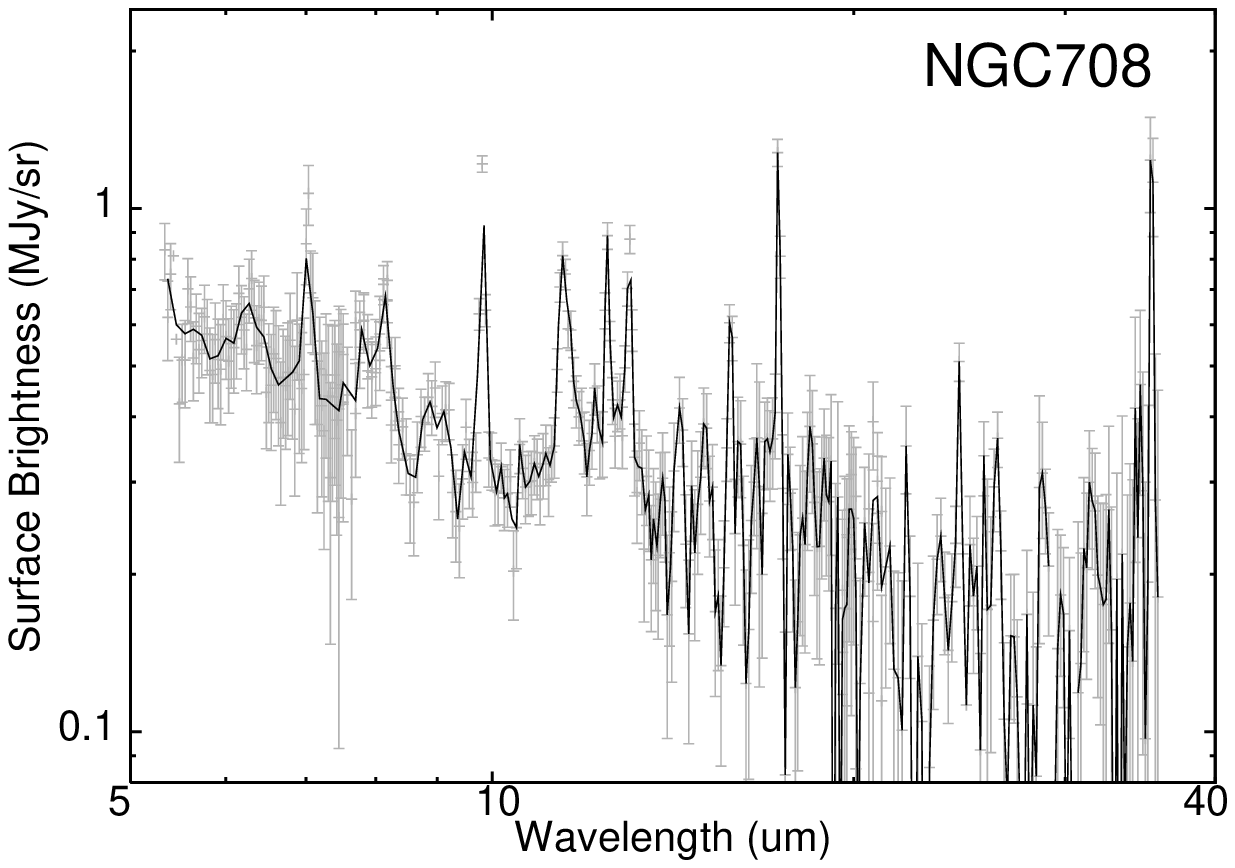}\plotone{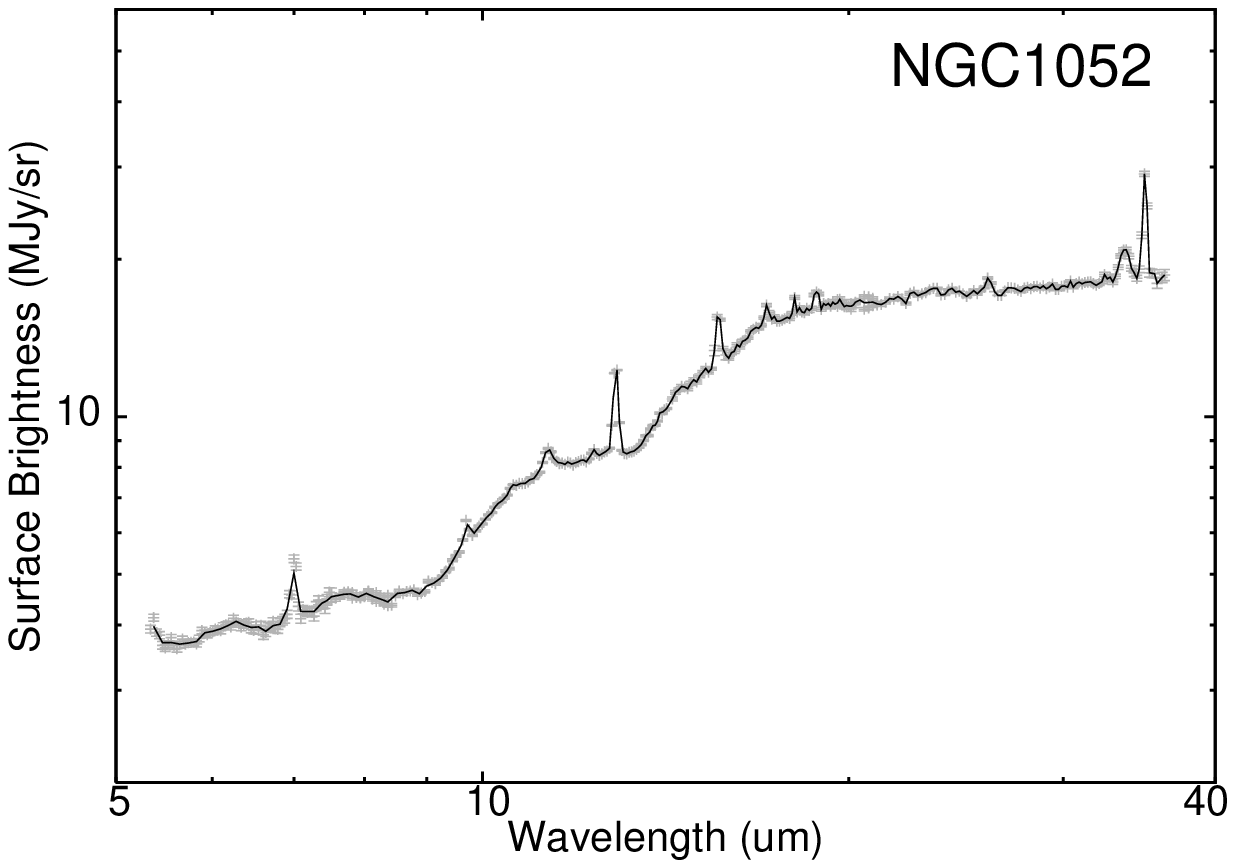}\\
\plotone{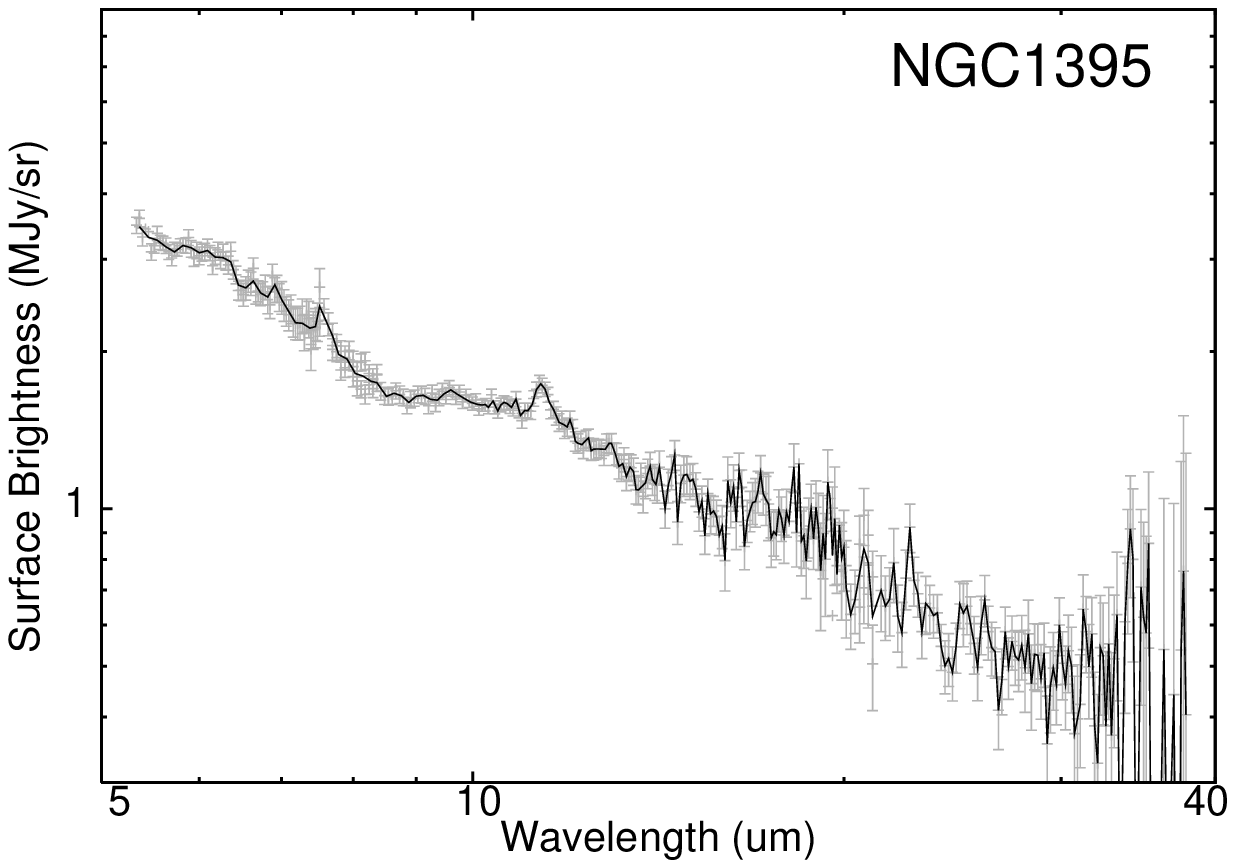}\plotone{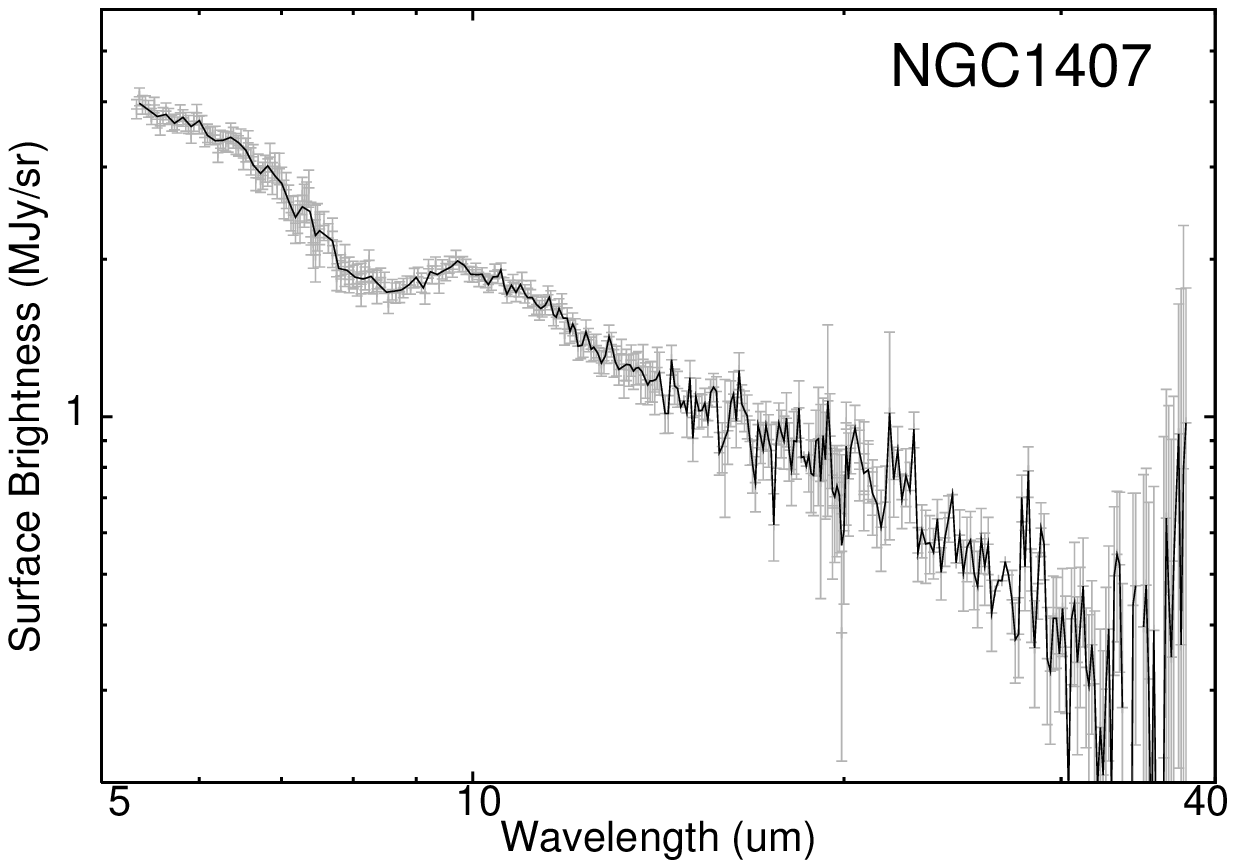}\\
\plotone{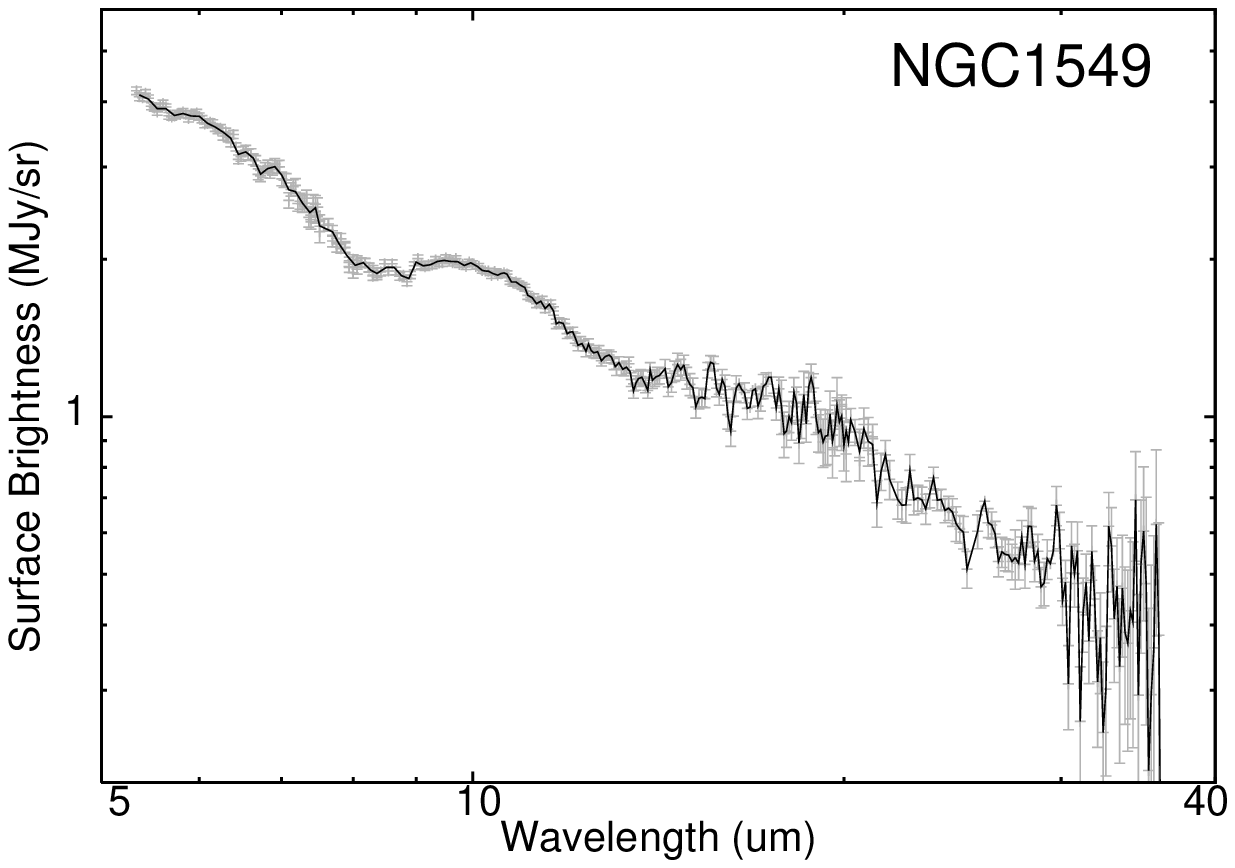}\plotone{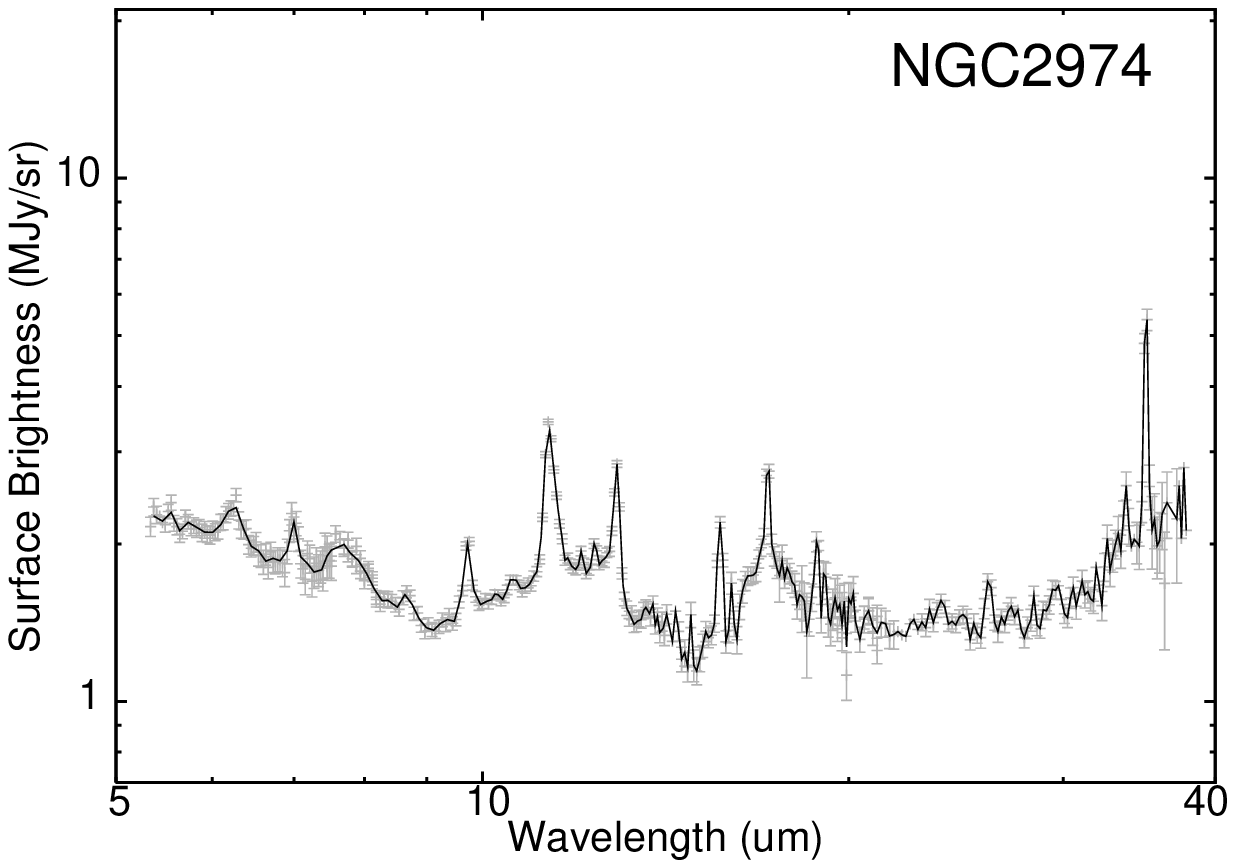}\\
\plotone{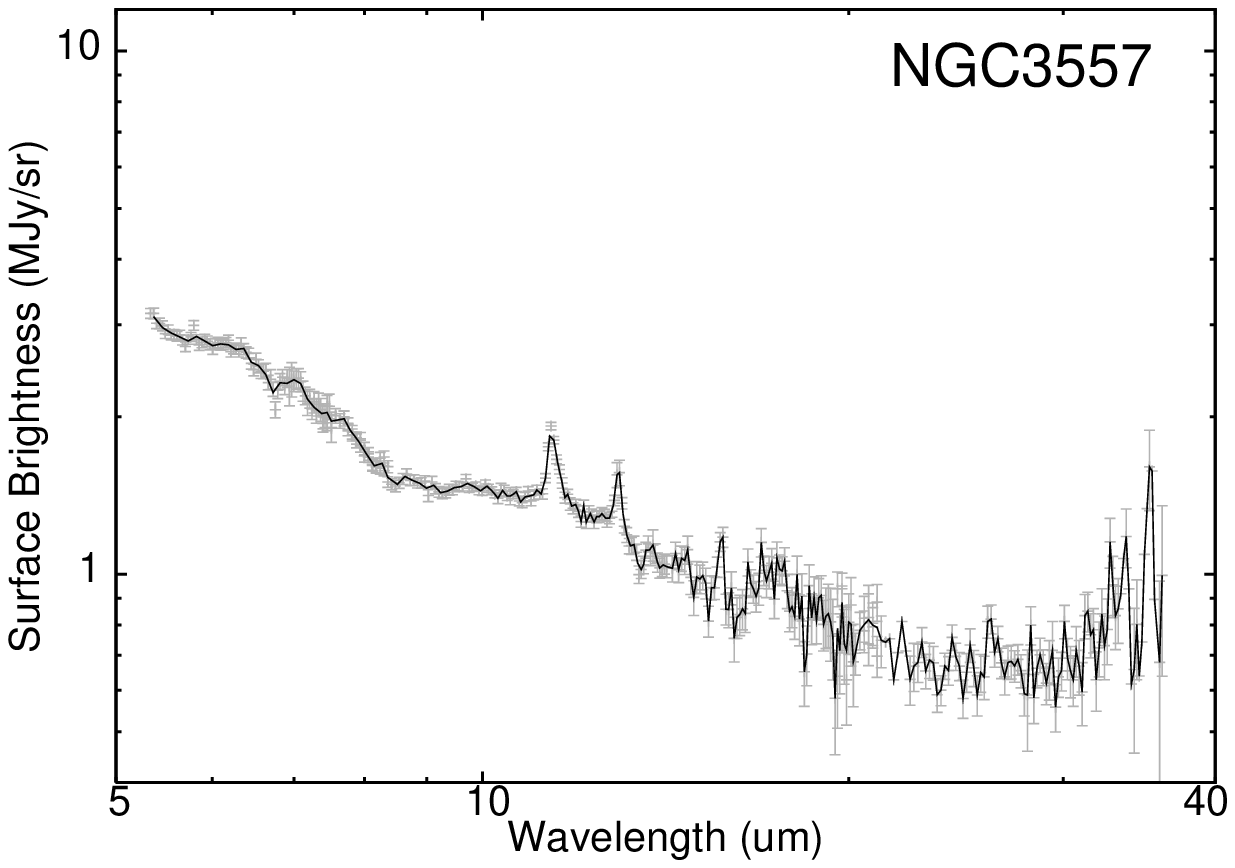}\plotone{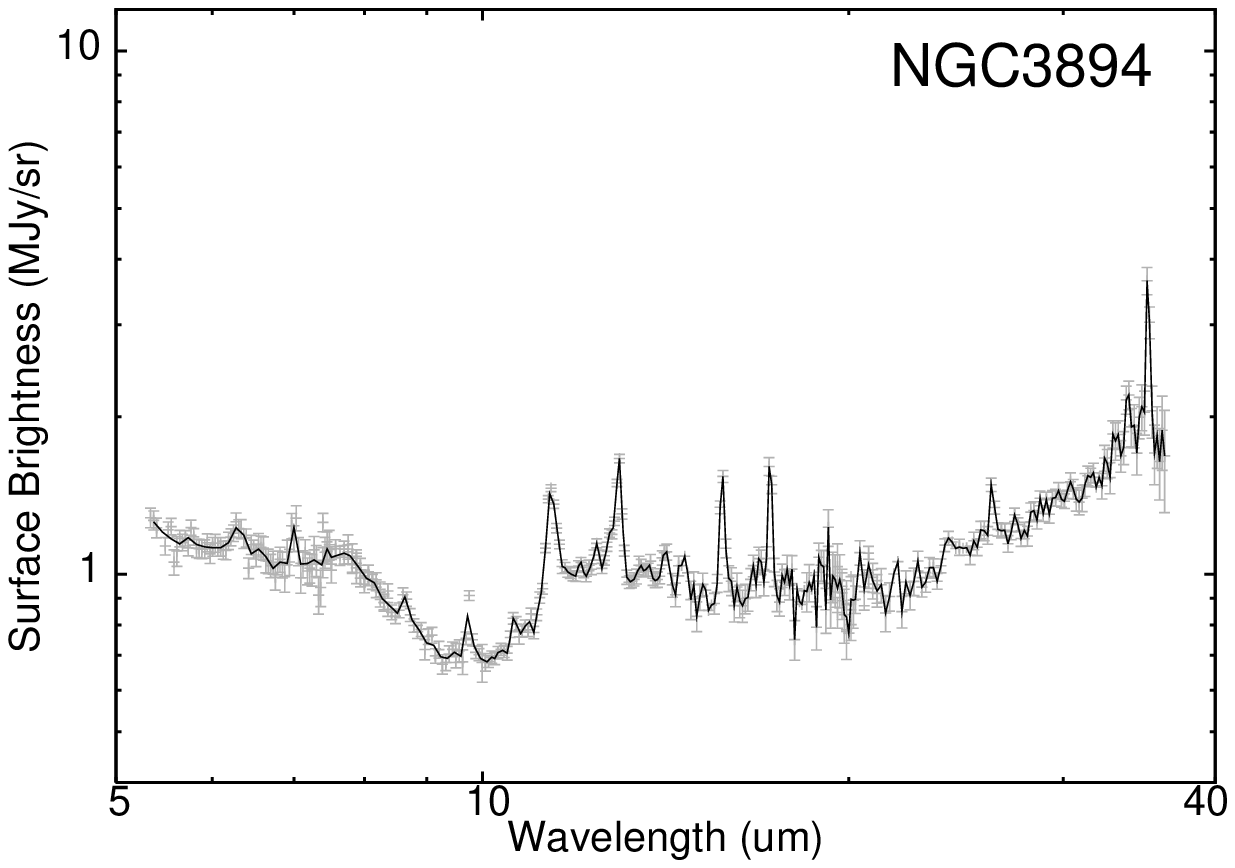}\\
\plotone{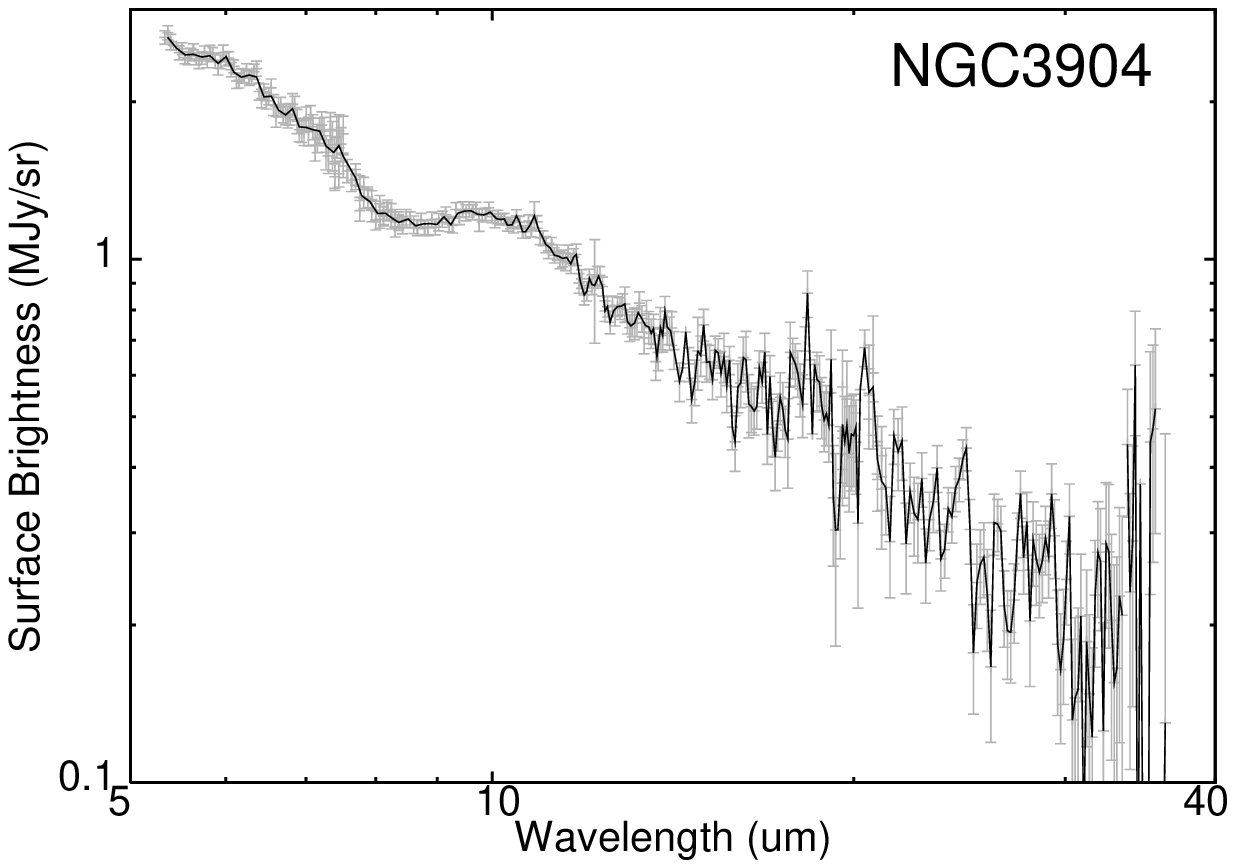}\plotone{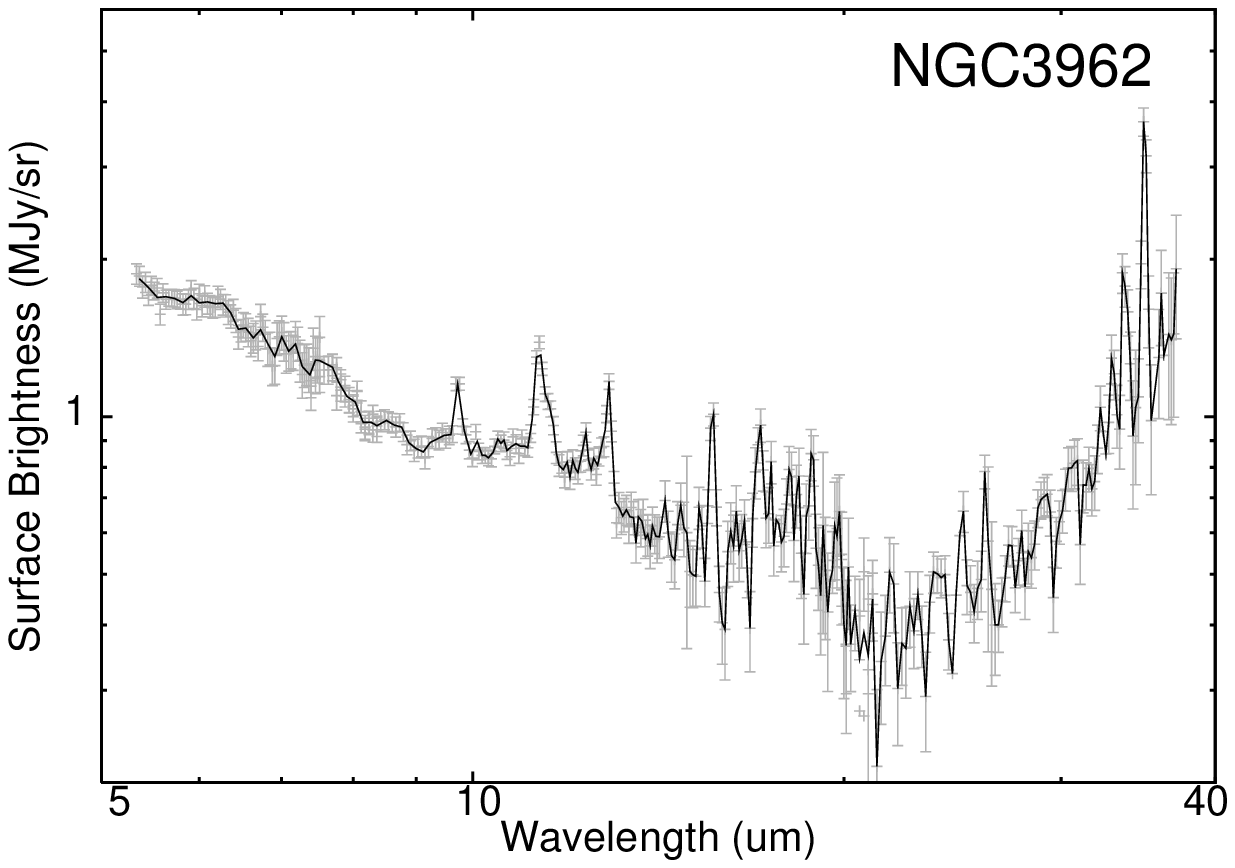}
\caption{Mid-IR spectra obtained with the IRS. The spectra were extracted and calibrated by using the SPICE (version 2.0.4) software package. Error bars are plotted together. Background spectra estimated from the observations of nearby blank skies have been subtracted.}
\end{figure}    
\clearpage
\begin{center}
\epsscale{.37}
\plotone{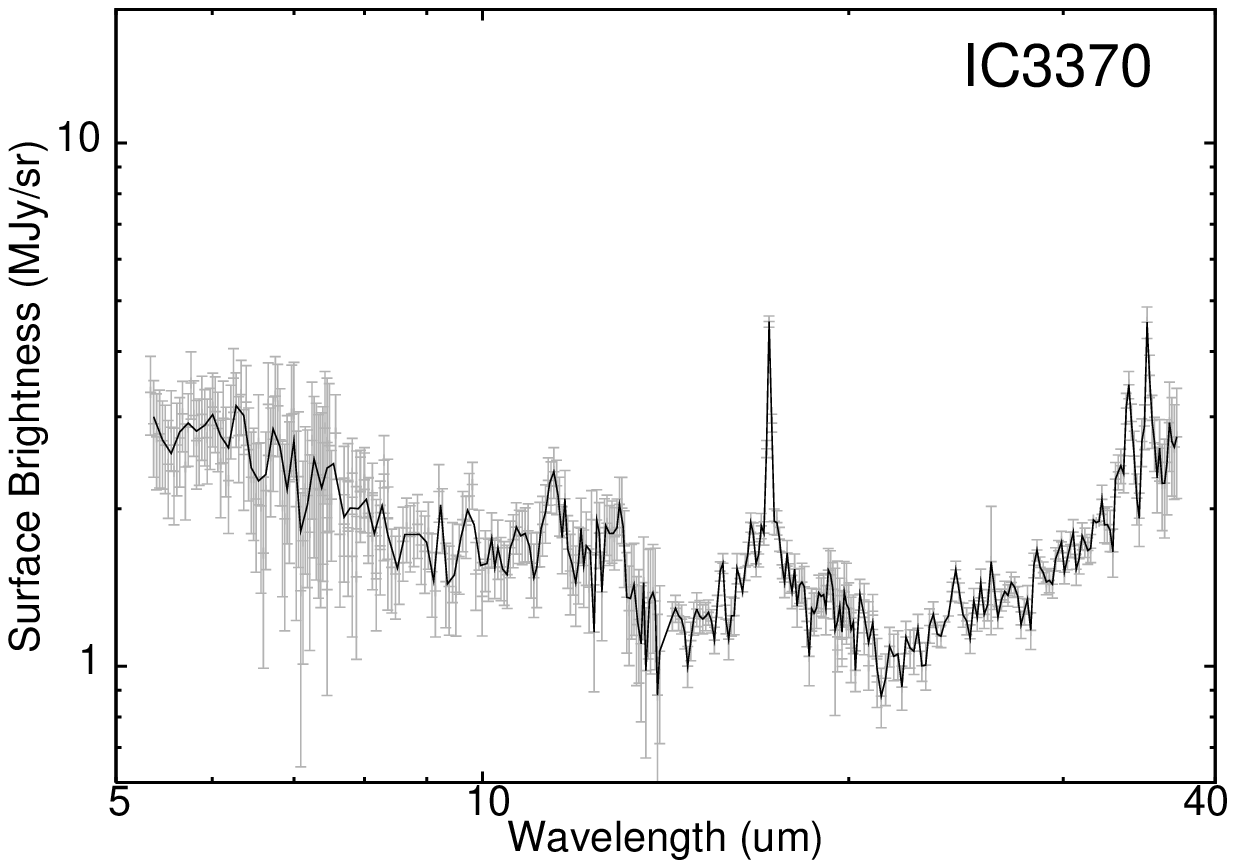}\plotone{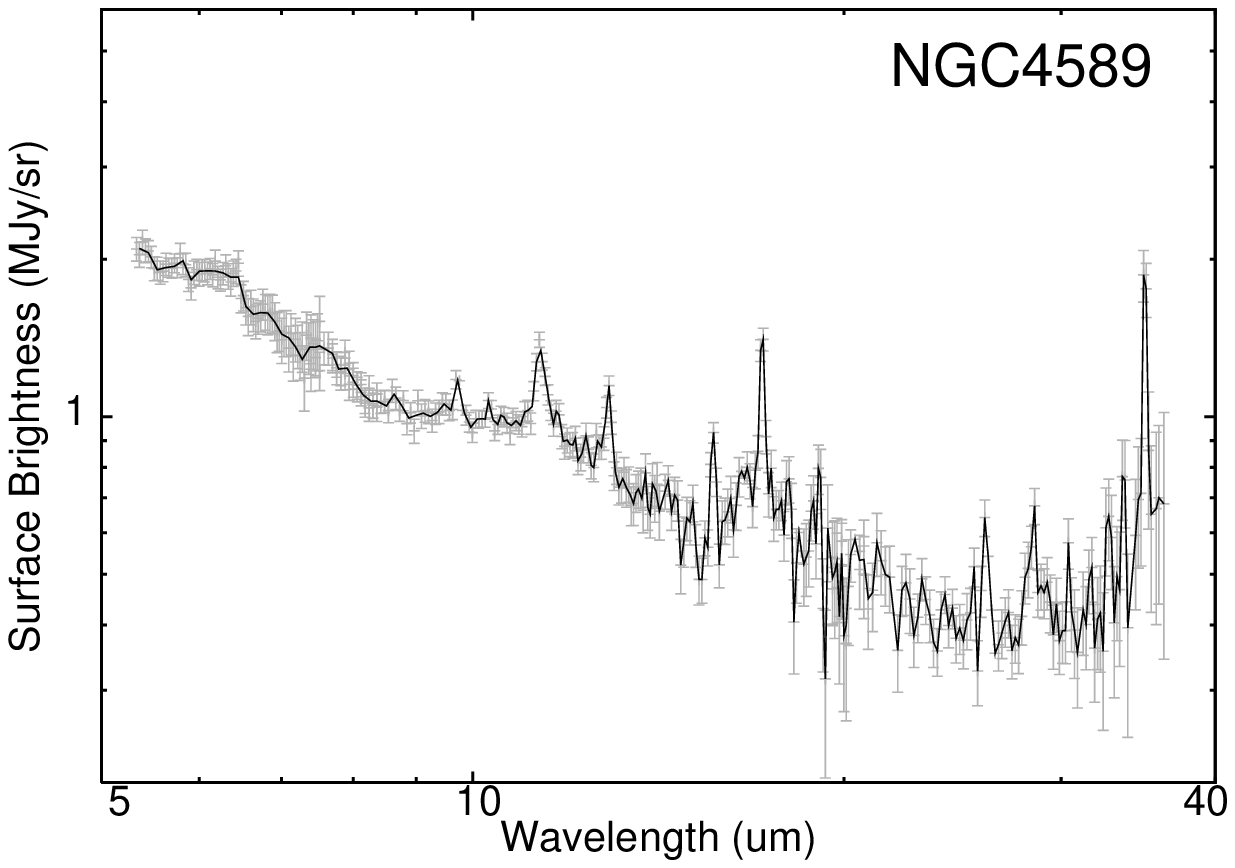}\\
\plotone{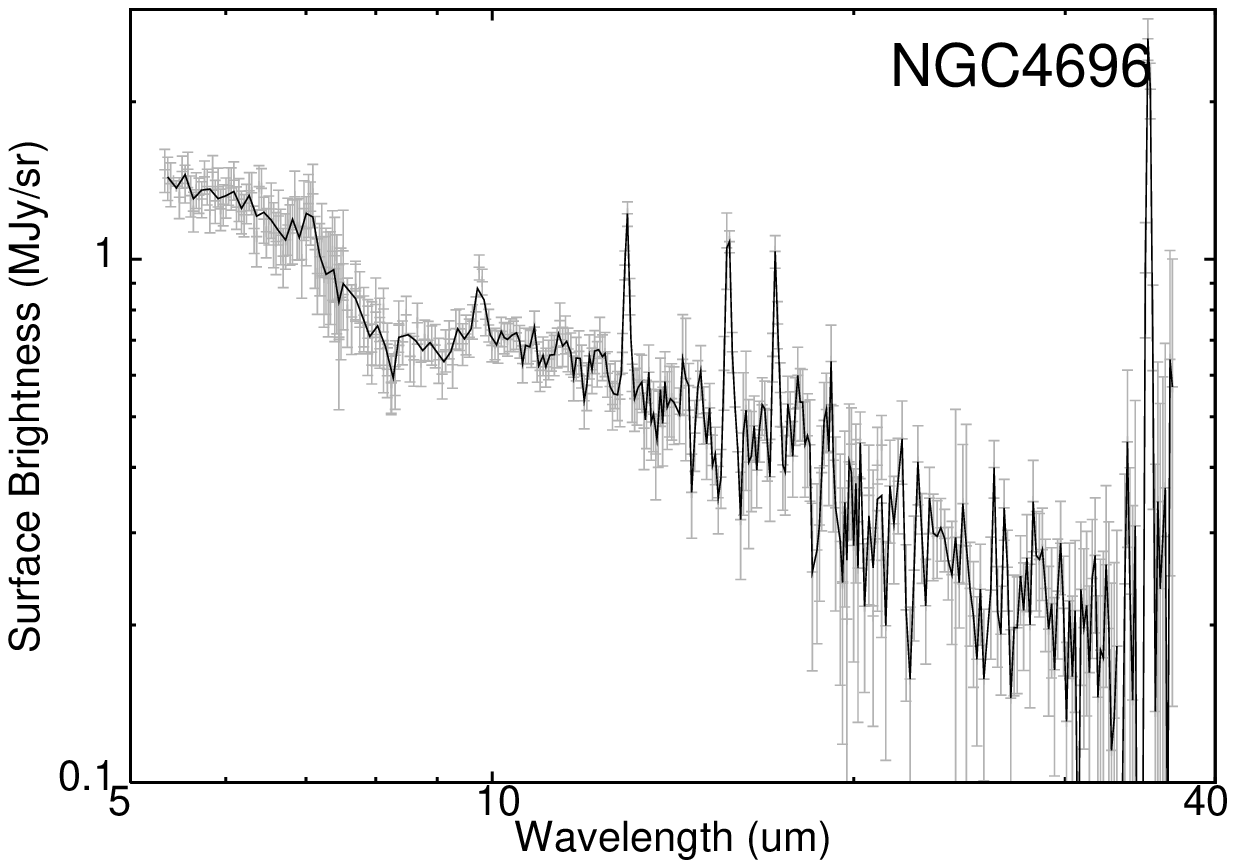}\plotone{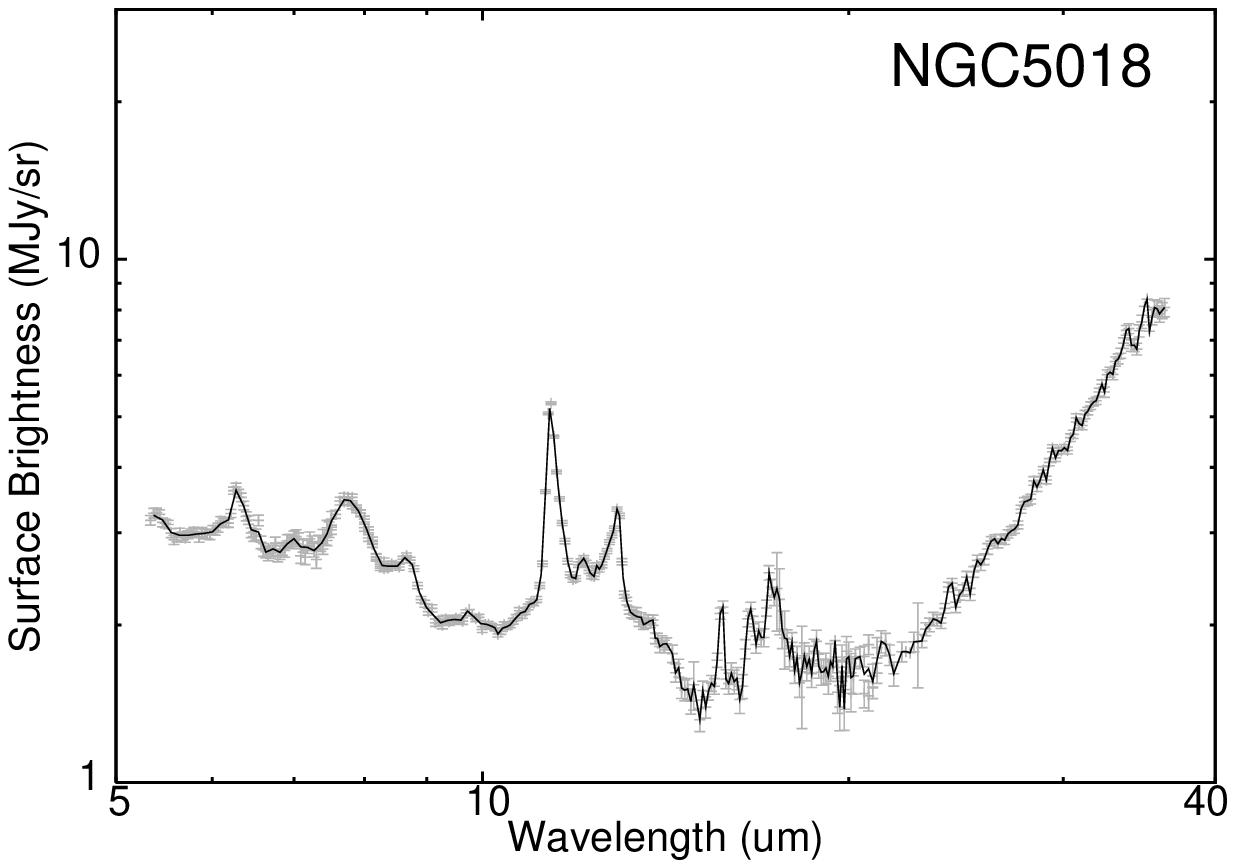}\\
\plotone{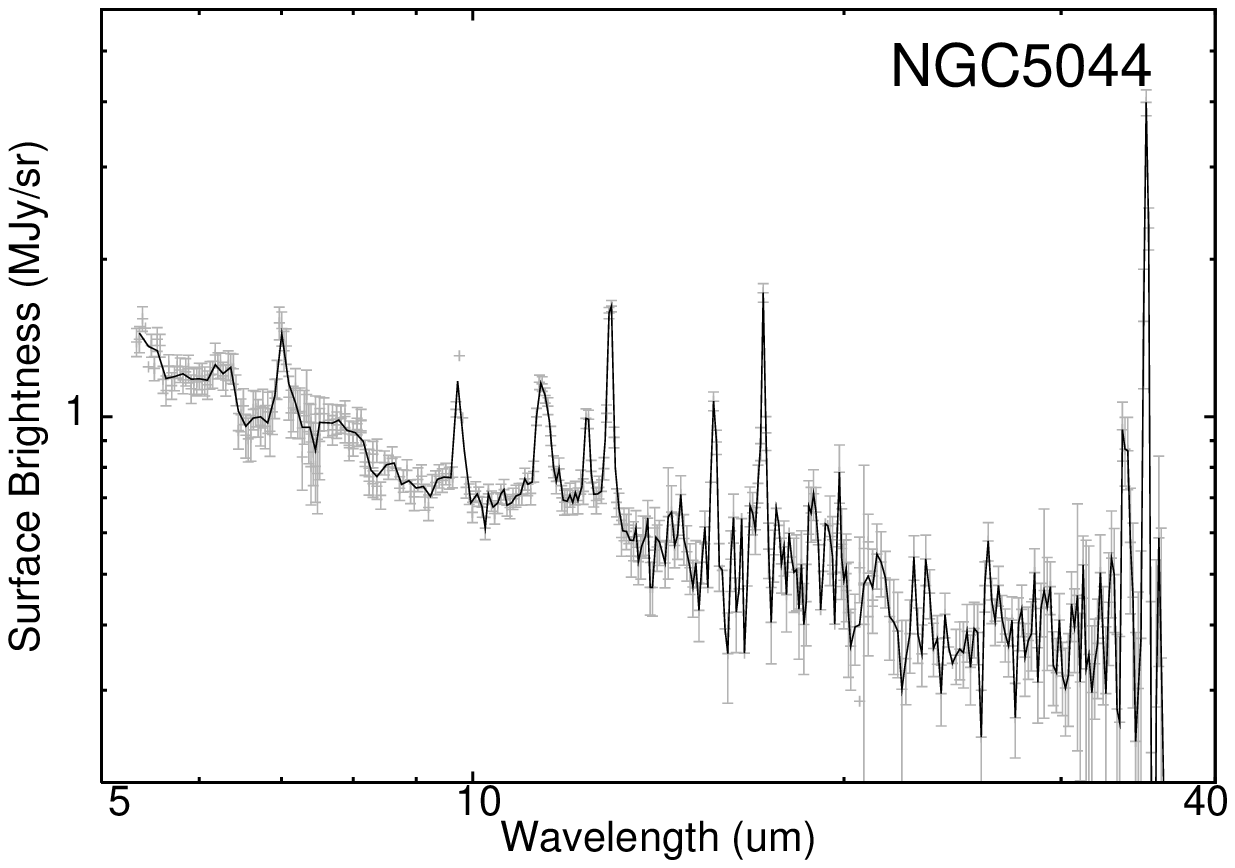}\plotone{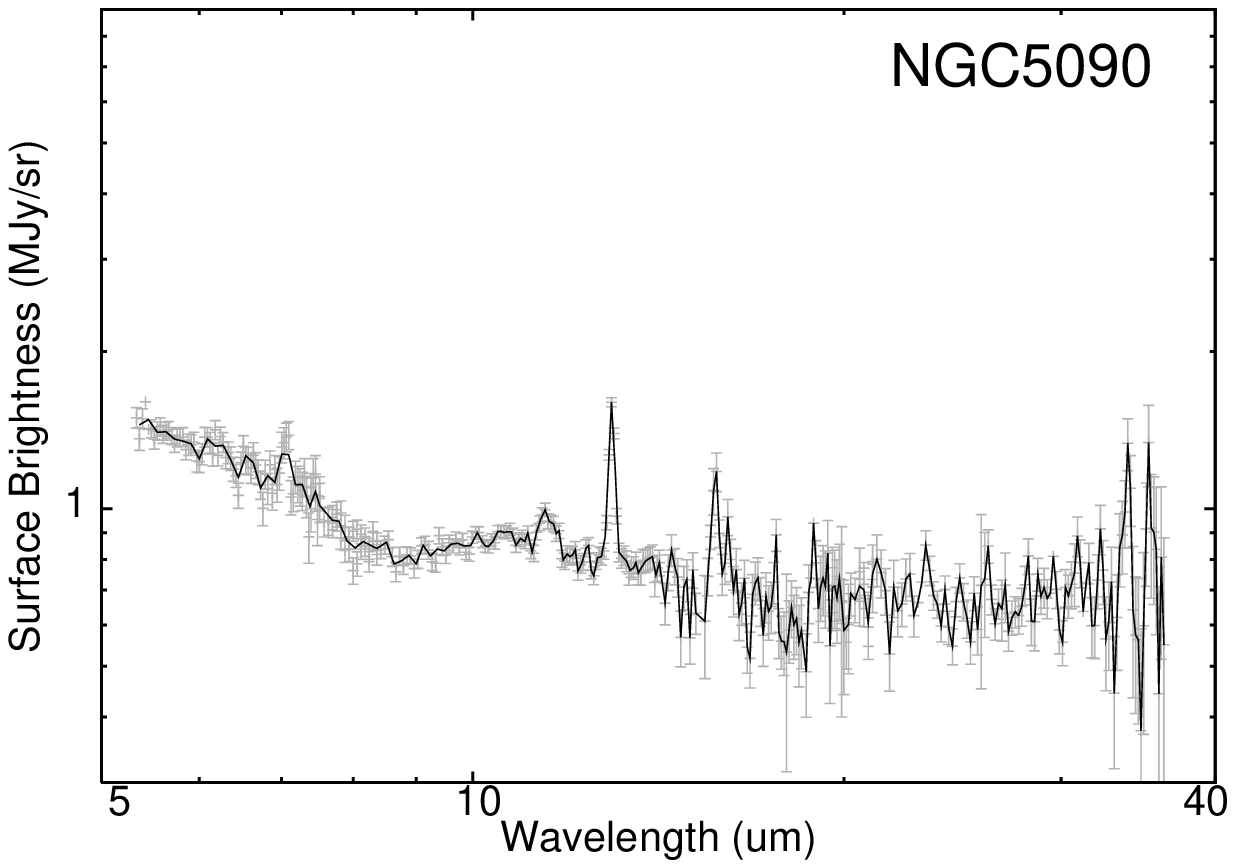}\\
\plotone{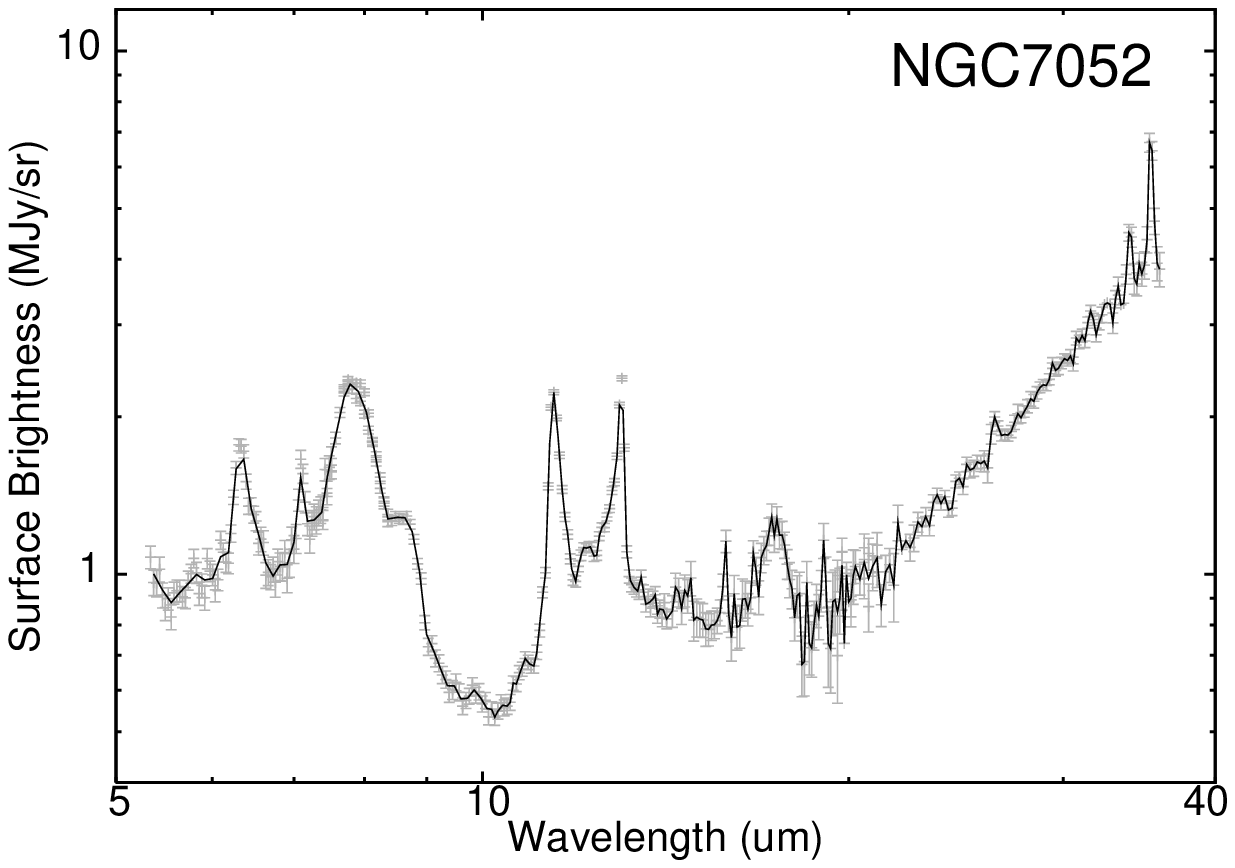}\plotone{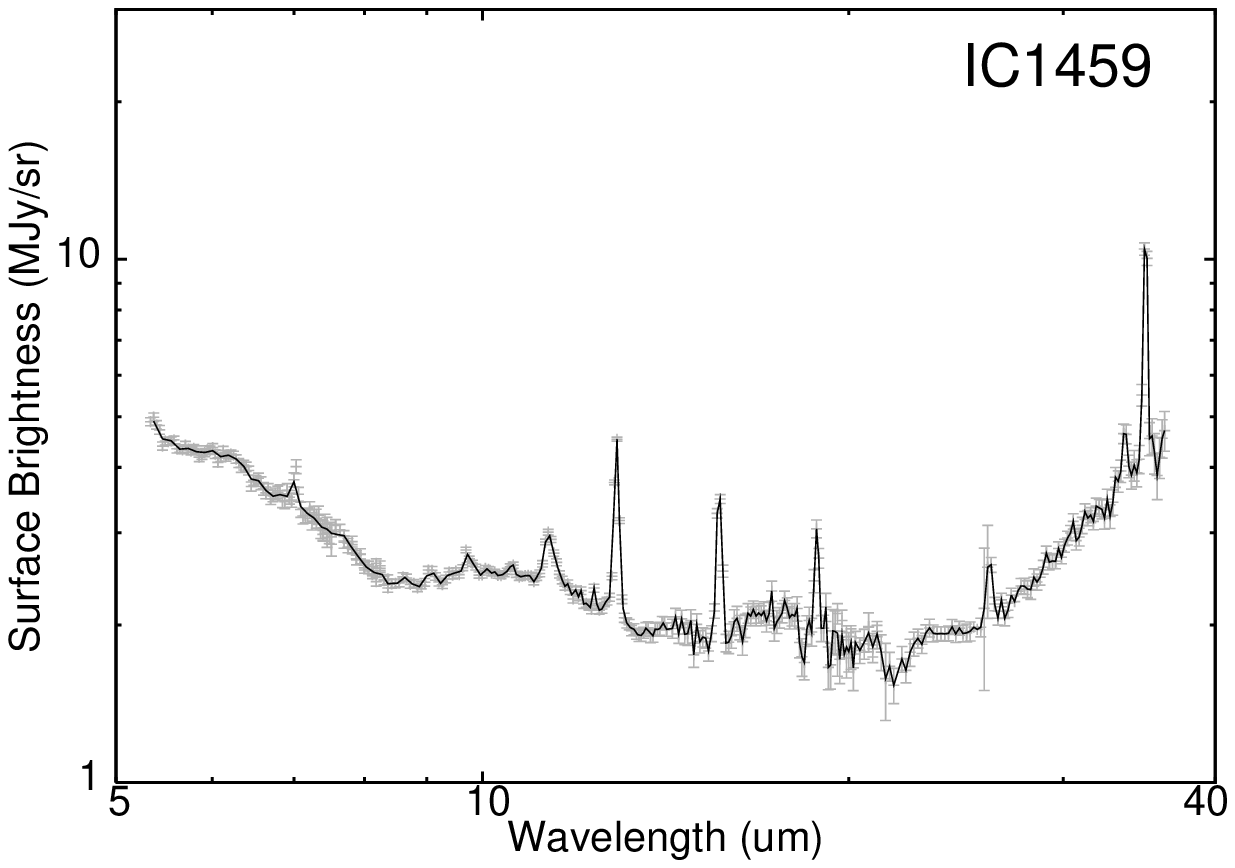}\\
{Fig.1. (continued).}
\end{center}
\clearpage

\begin{figure}
\epsscale{.37}
\plotone{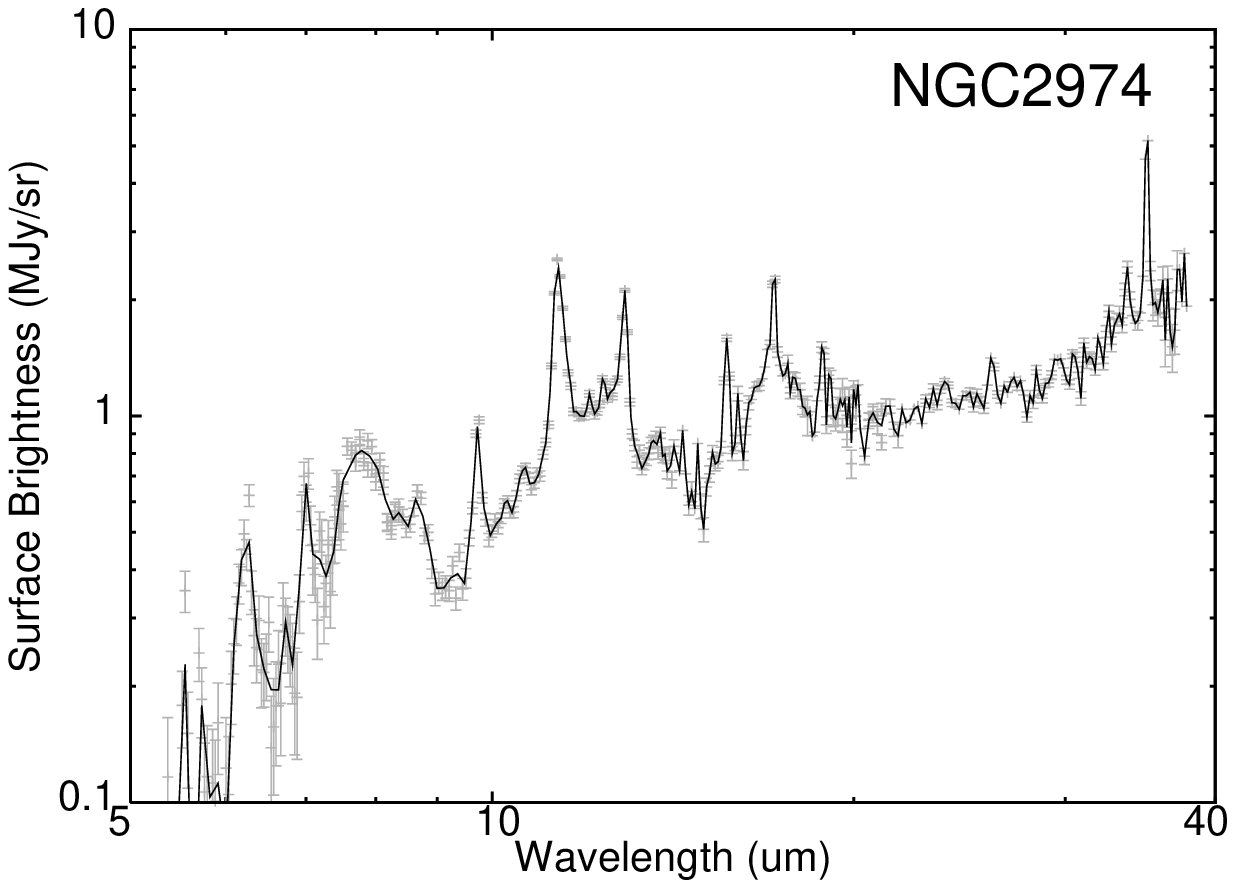}\plotone{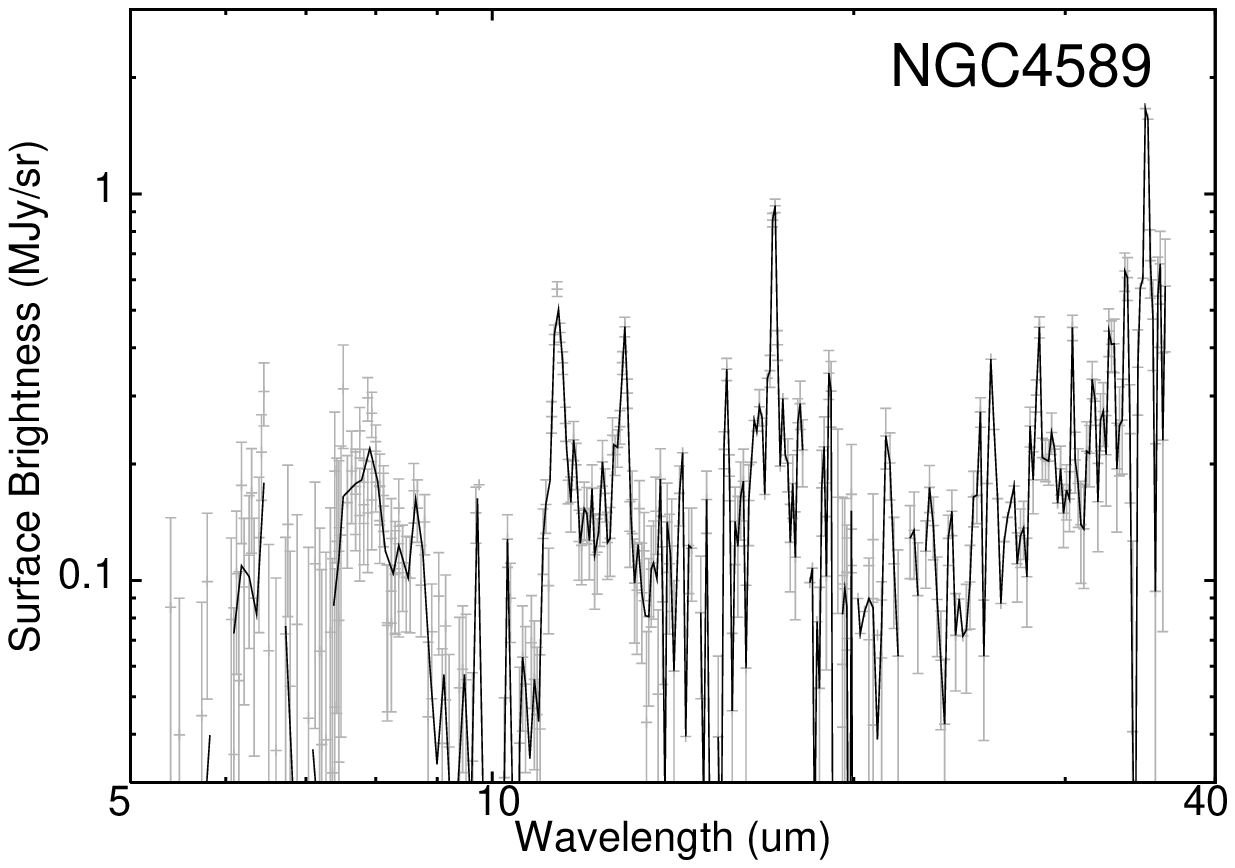}\\
\plotone{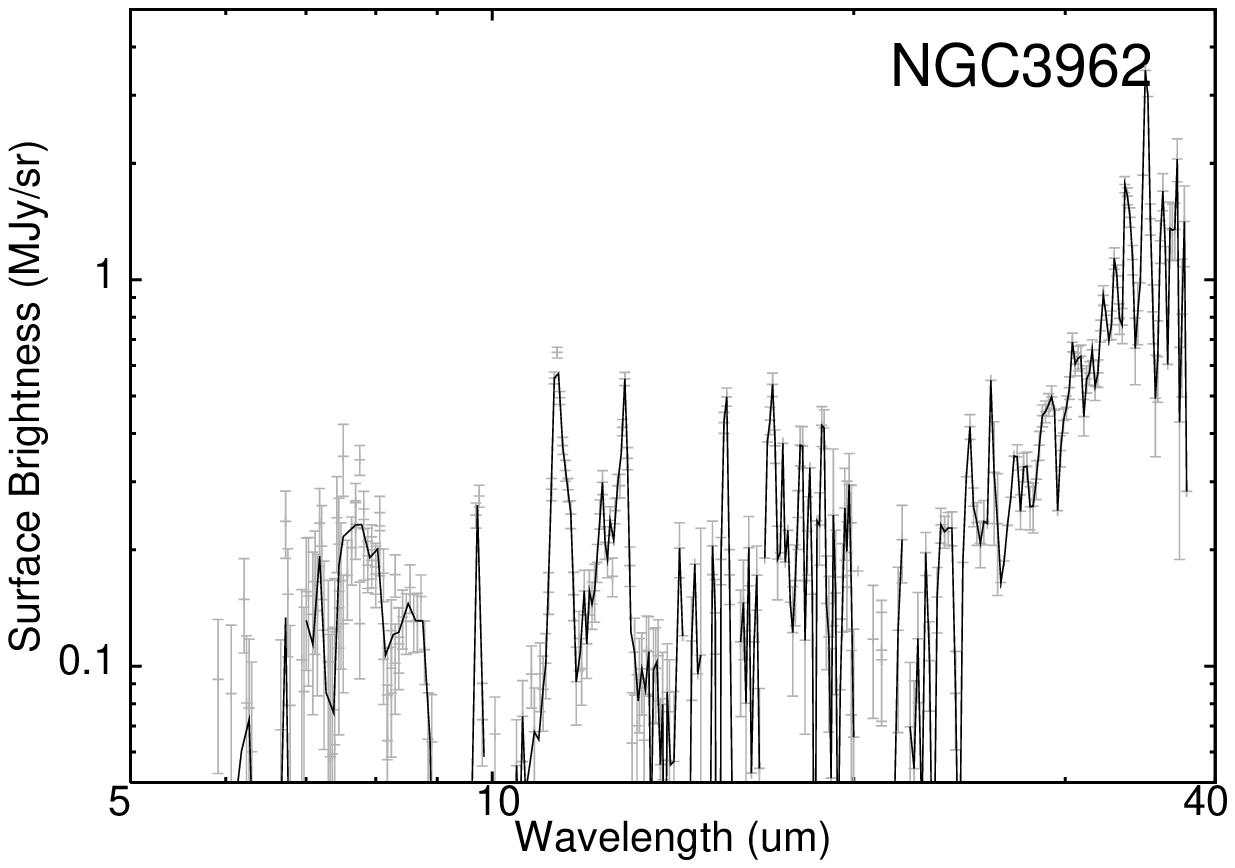}\plotone{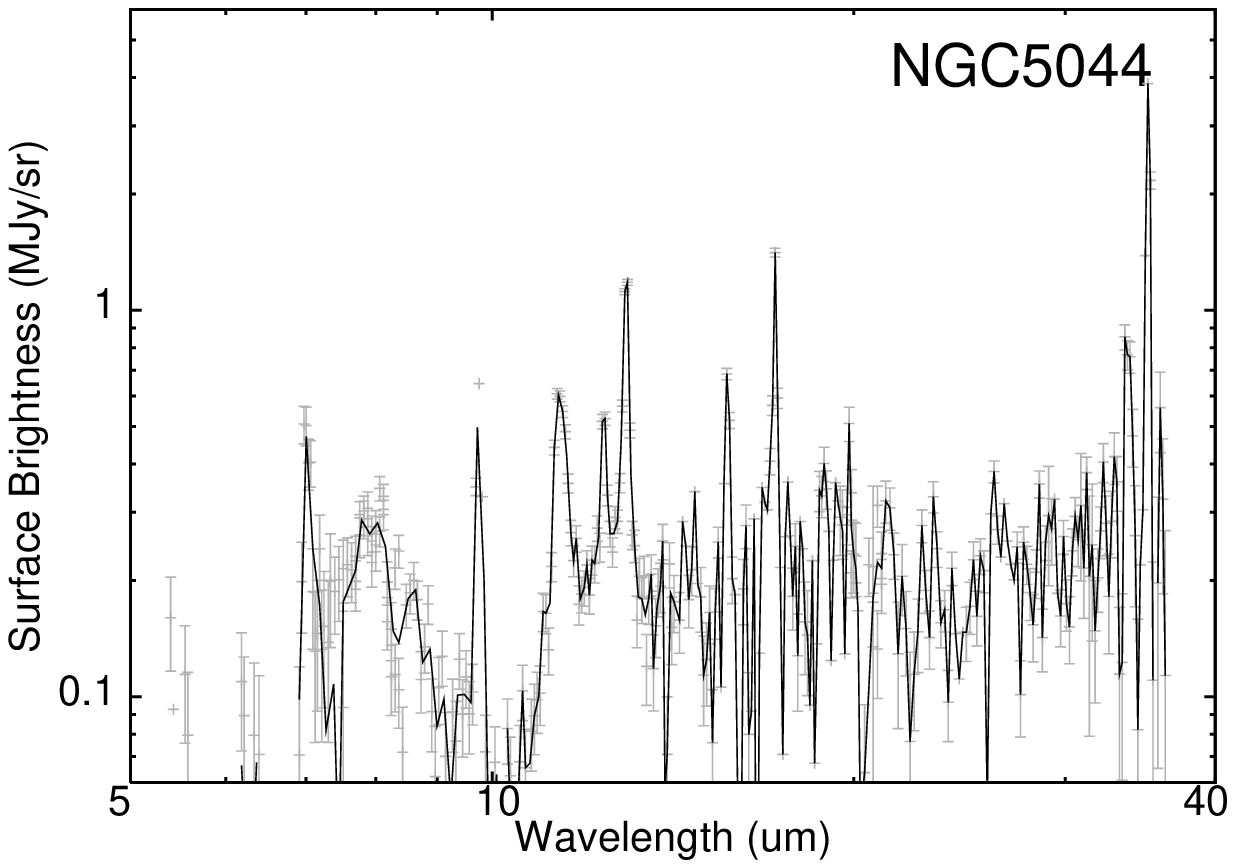}\\
\epsscale{.45}
\plotone{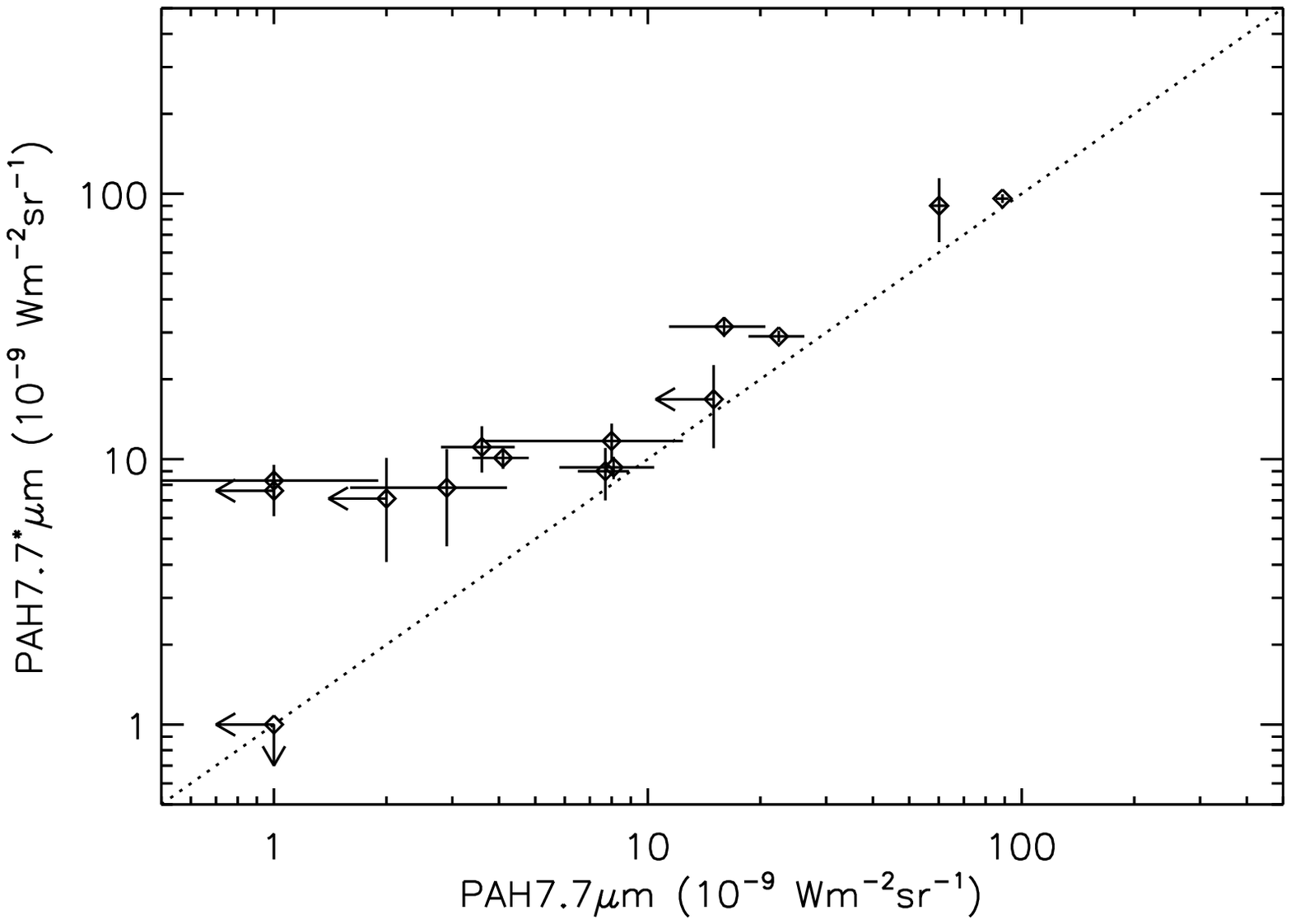}
\caption{Stellar-component-subtracted spectra for the representative elliptical galaxies (see text). The bottom panel shows the comparison of the PAH 7.7$\mu$m feature strengths obtained by fitting the spectra before (PAH 7.7) and after (PAH 7.7$^*$) the subtraction of the stellar component, together with a slope unity line.}
\end{figure}

\begin{figure}
\epsscale{.45}
\plotone{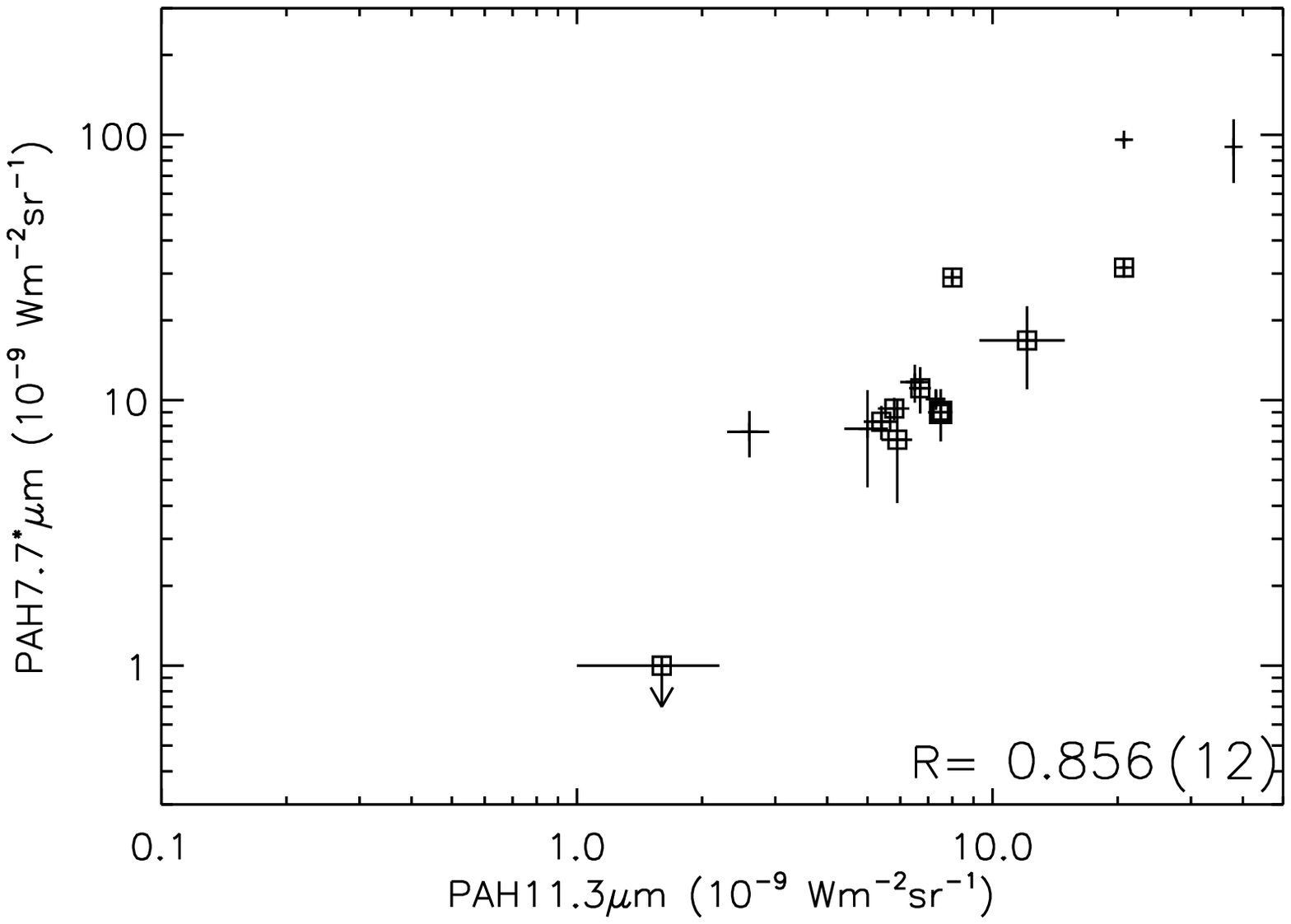}\plotone{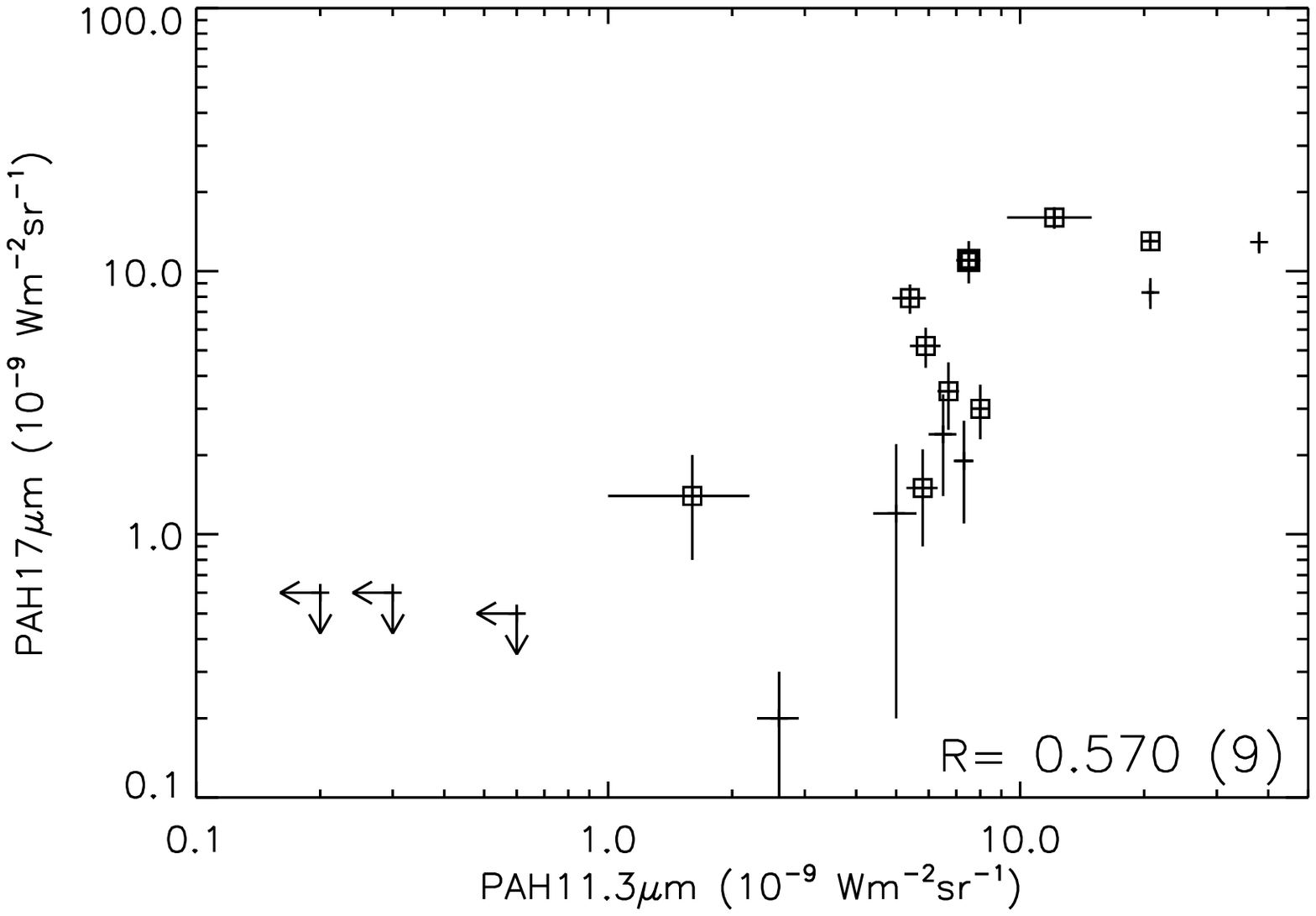}\\
\plotone{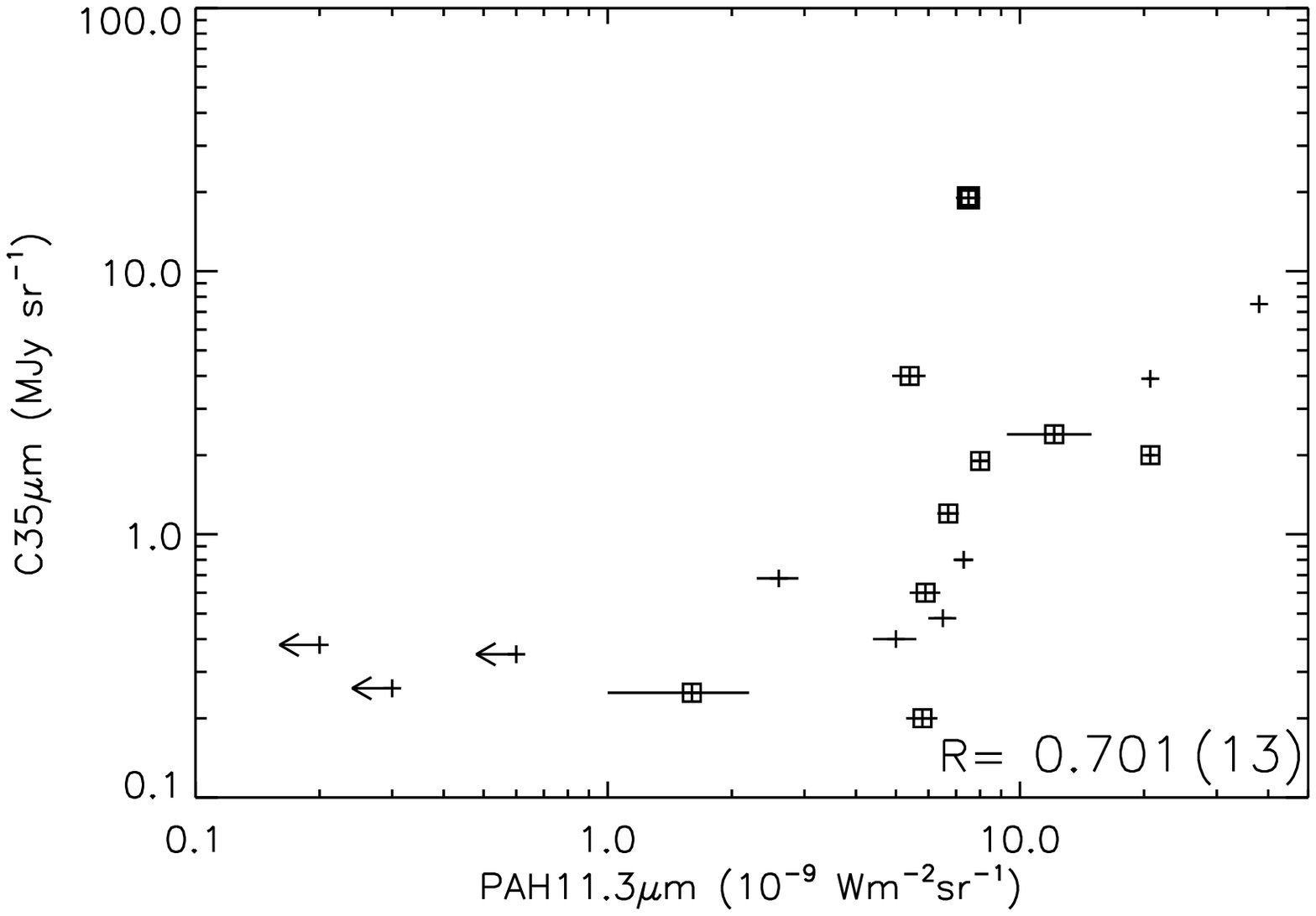}\plotone{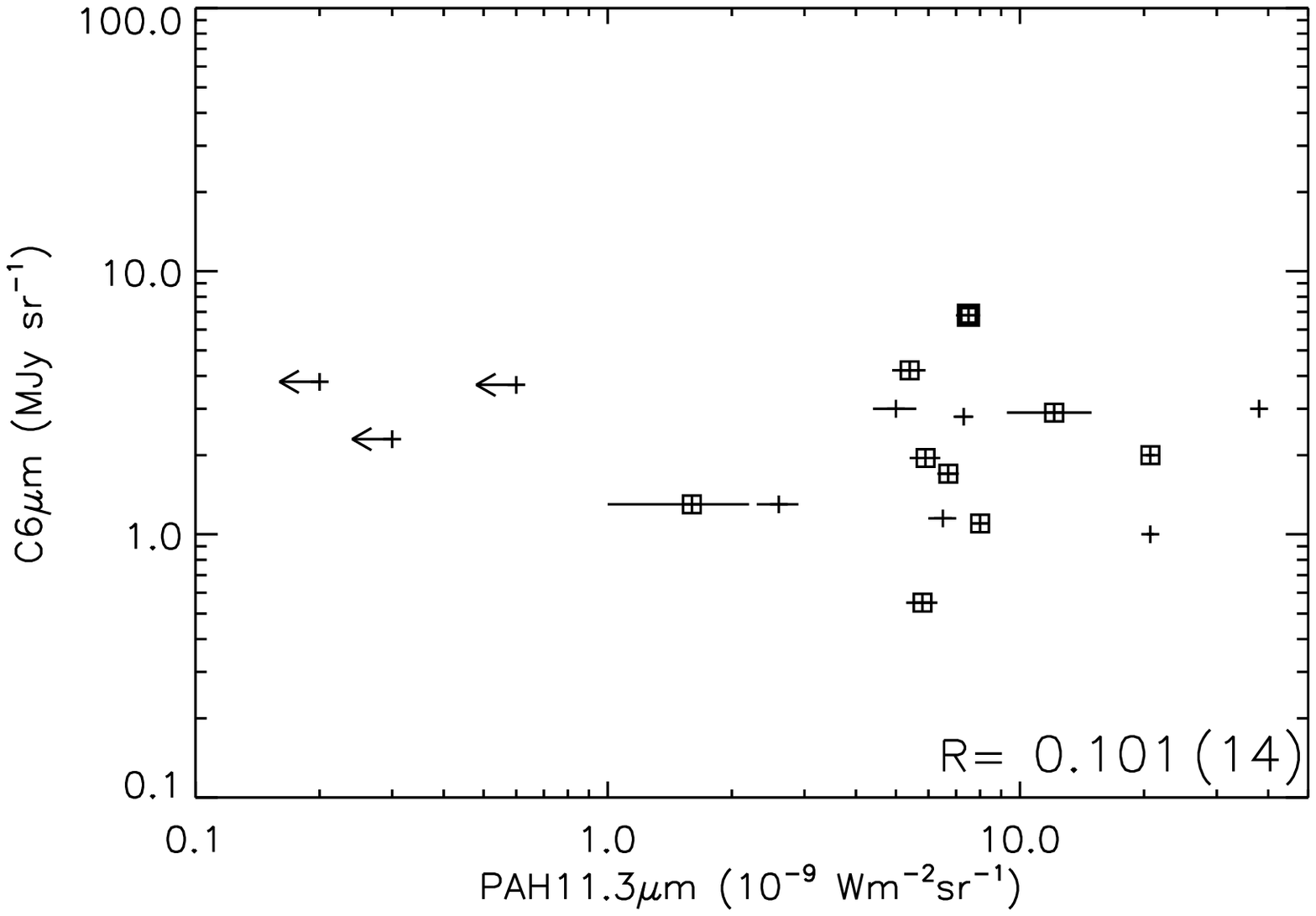}\\
\plotone{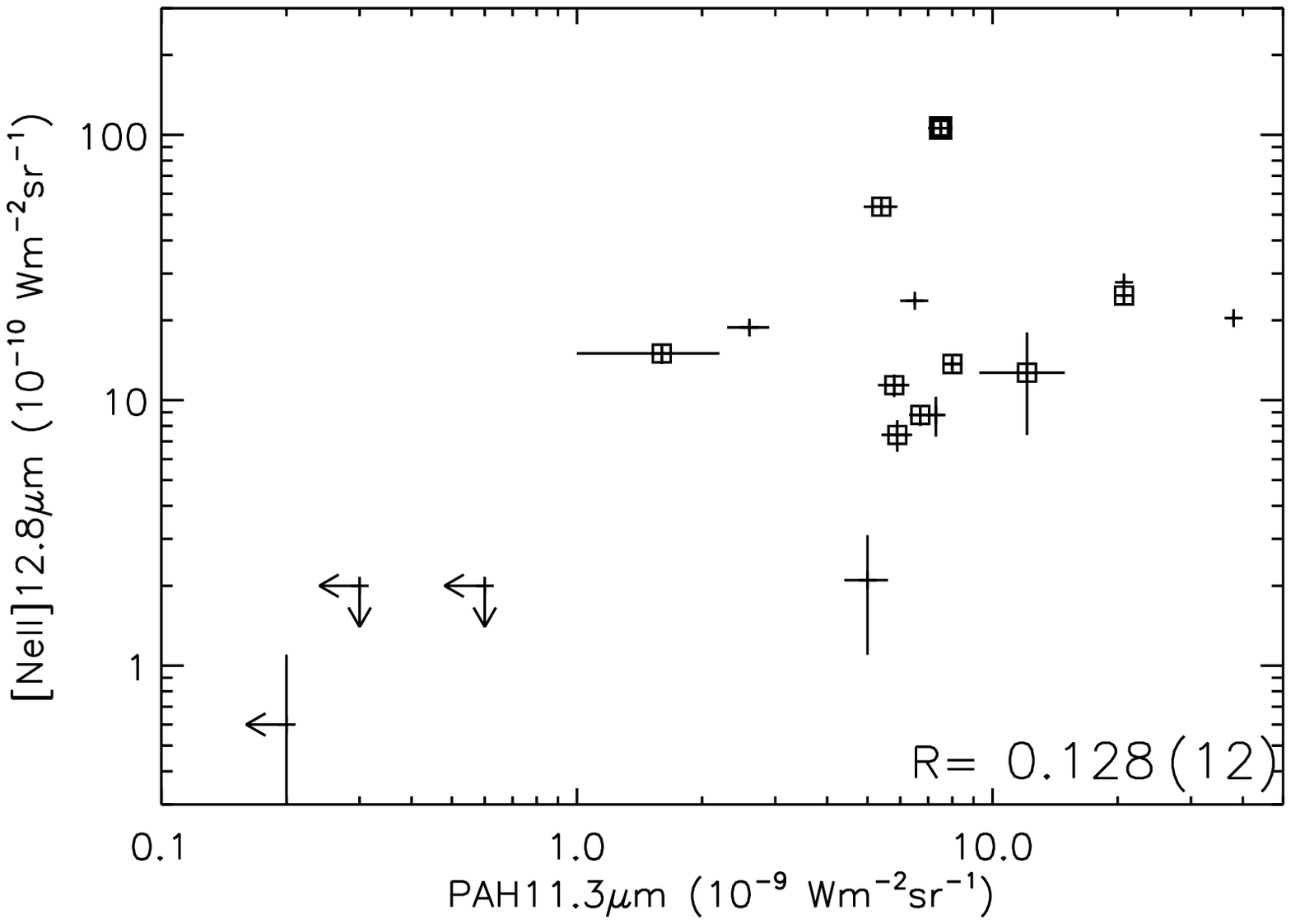}\plotone{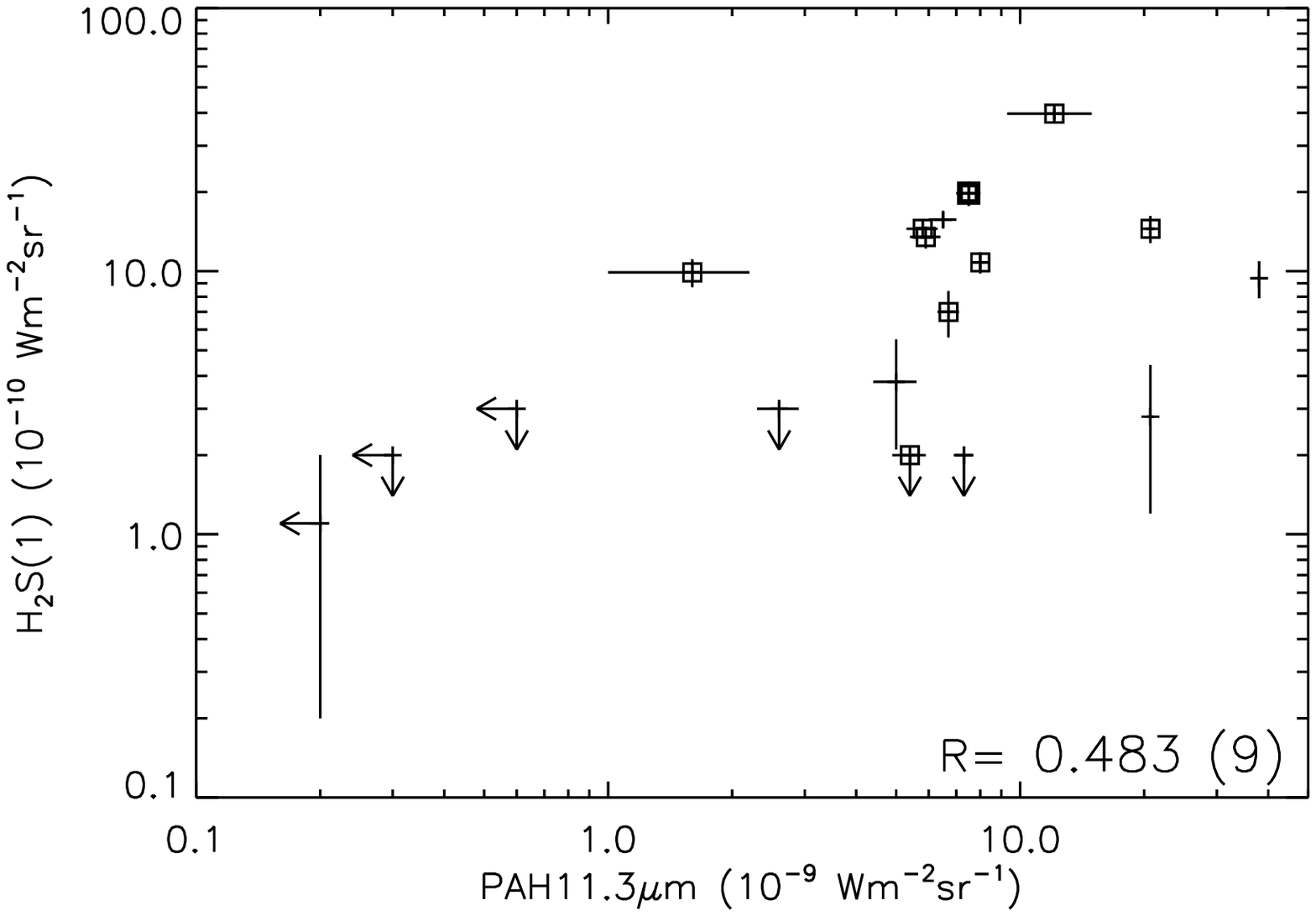}
\caption{Variations of the strength of the PAH 11.3 $\mu$m feature with respect to those of other spectral components. The PAH 7.7$^*$ denotes the strength of the PAH 7.7$\mu$m feature obtained from the stellar-component-subtracted spectra (Fig2). $R$ in the figure is a linear-correlation coefficient for the correponding plot, where the number in the parentheses corresponds to that of the data points (S/N$>$3) used in the calculation. The box indicates that the galaxy have a LLAGN. The thick box corresponds to the plot for NGC~1052 (see text). }
\end{figure}    
\clearpage

\begin{figure}
\epsscale{.45}
\plotone{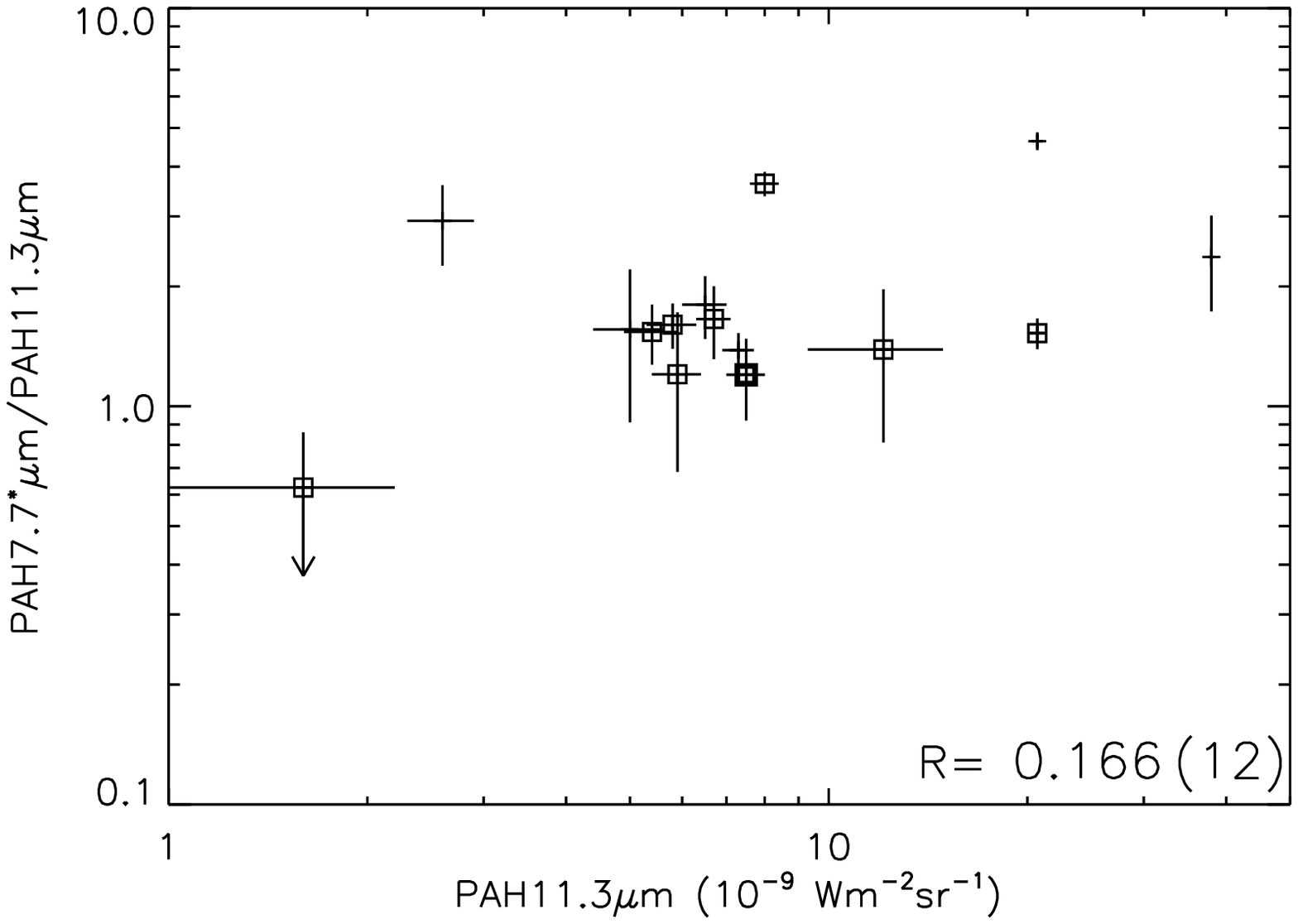}\\
\plotone{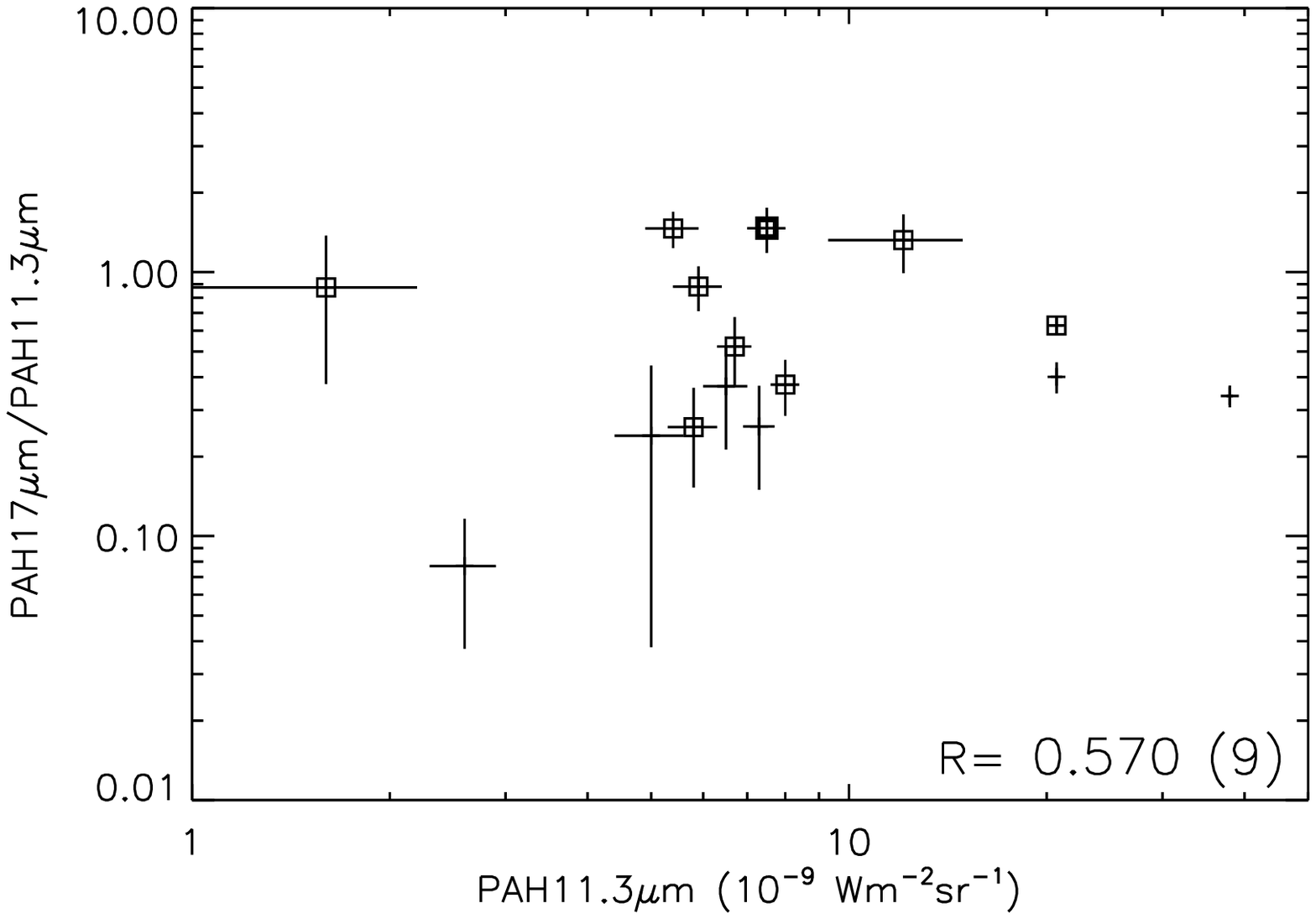}
\caption{PAH 7.7$^*$/11.3 and PAH 17/11.3 ratios plotted against the strength of the PAH 11.3 $\mu$m feature. The PAH 7.7$^*$ denotes the strength of the PAH 7.7$\mu$m feature obtained from the stellar-component-subtracted spectra (Fig2).}
\end{figure}    
\clearpage

\begin{figure}
\epsscale{.45}
\plotone{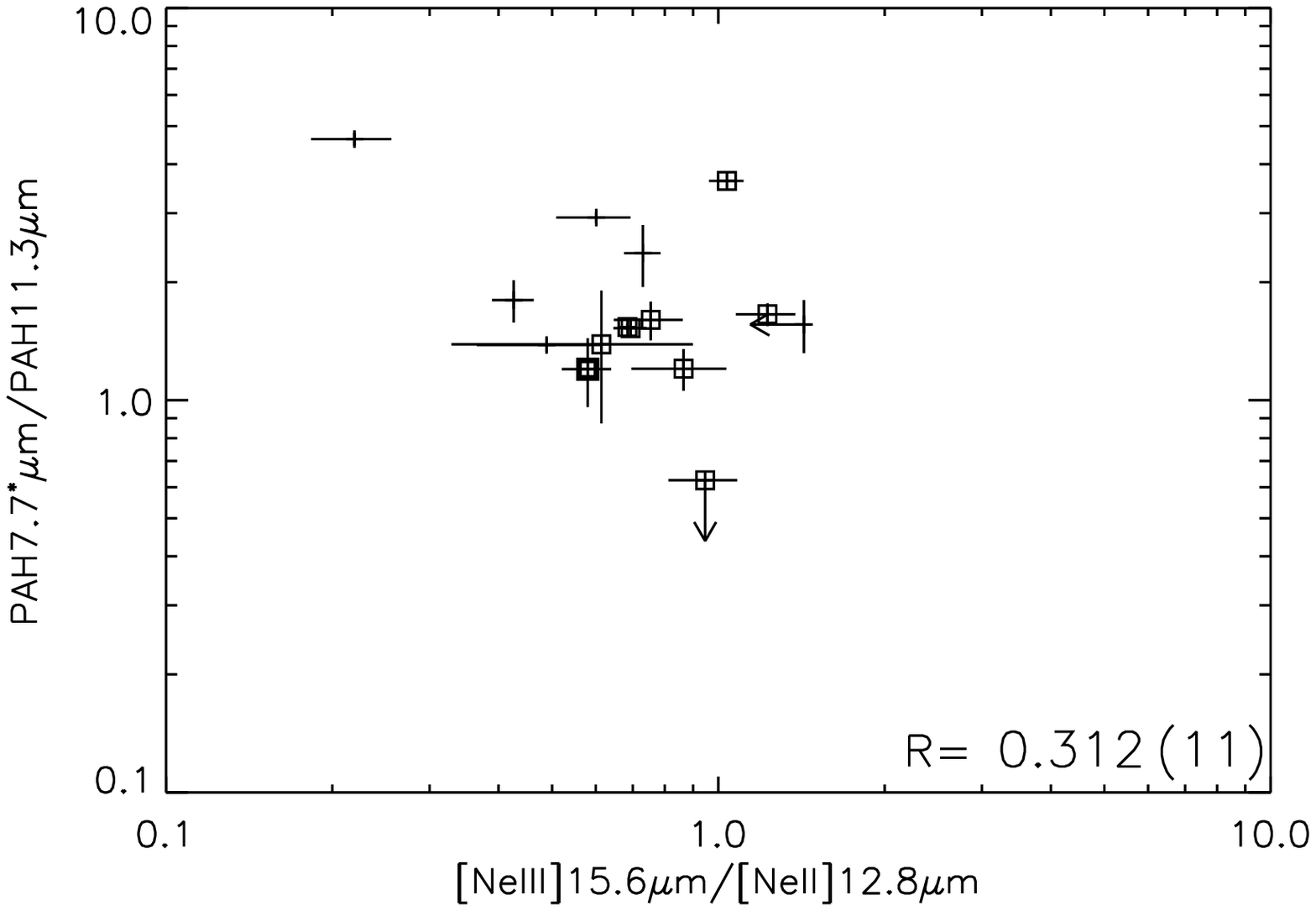}\plotone{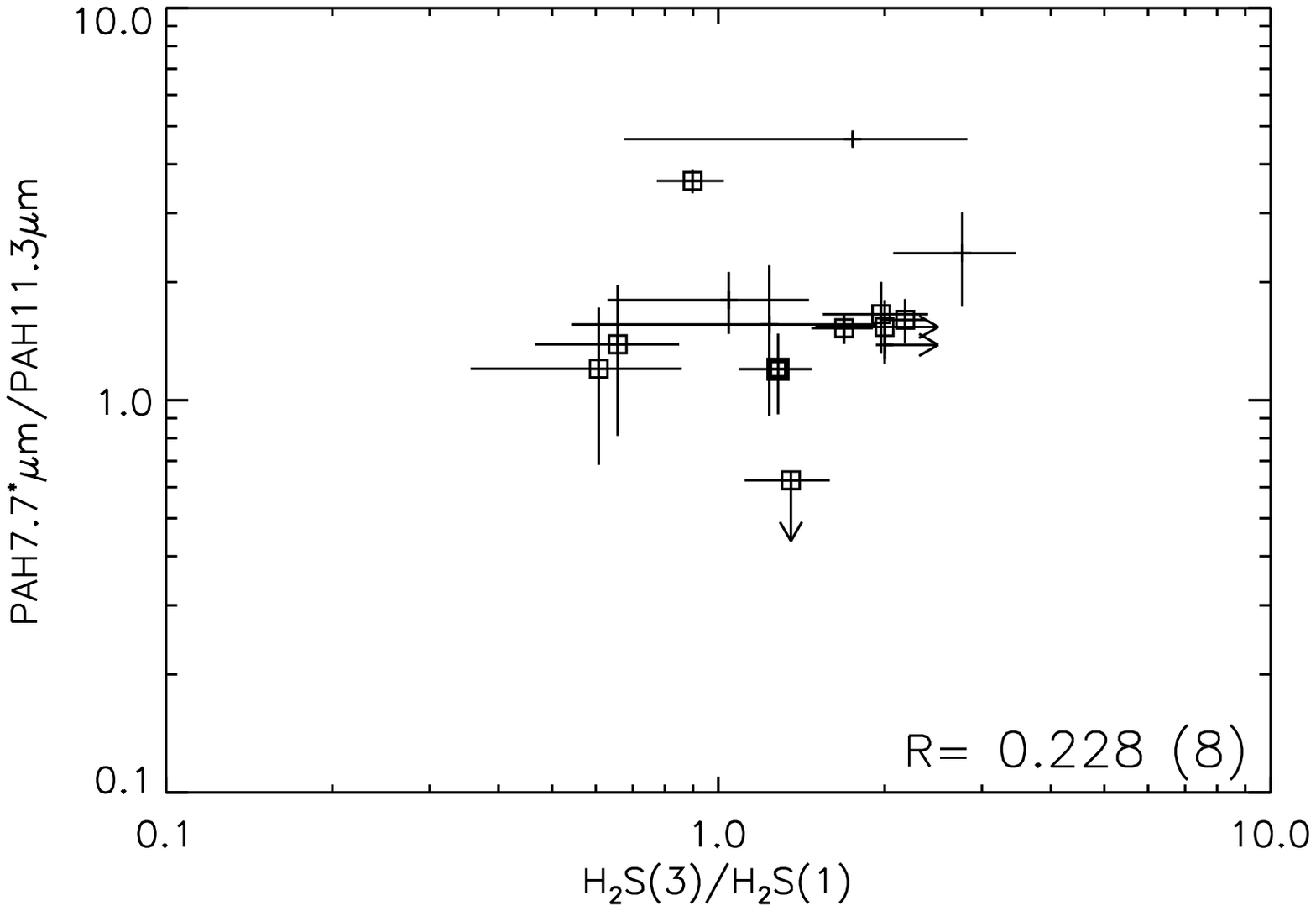}\\
\plotone{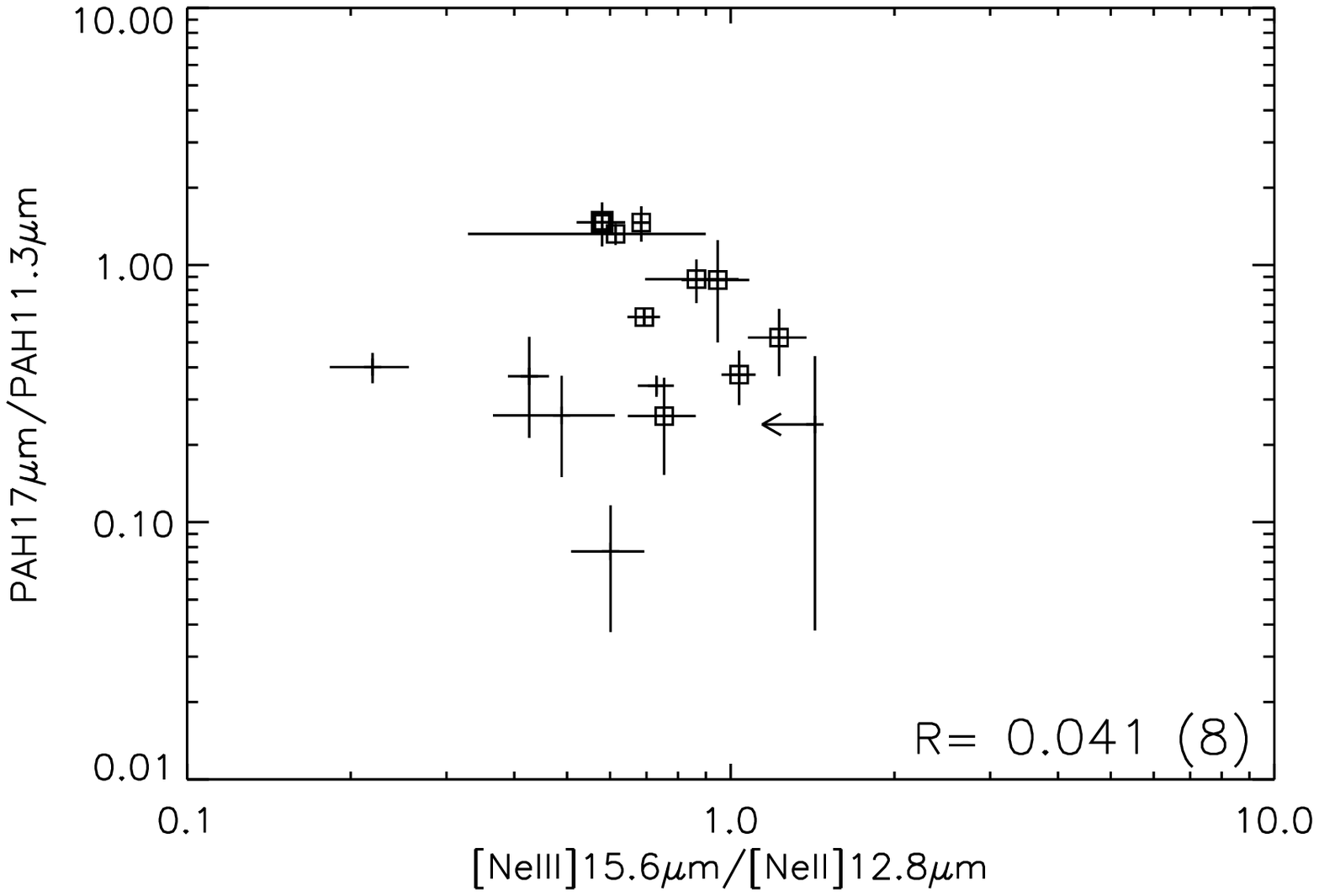}\plotone{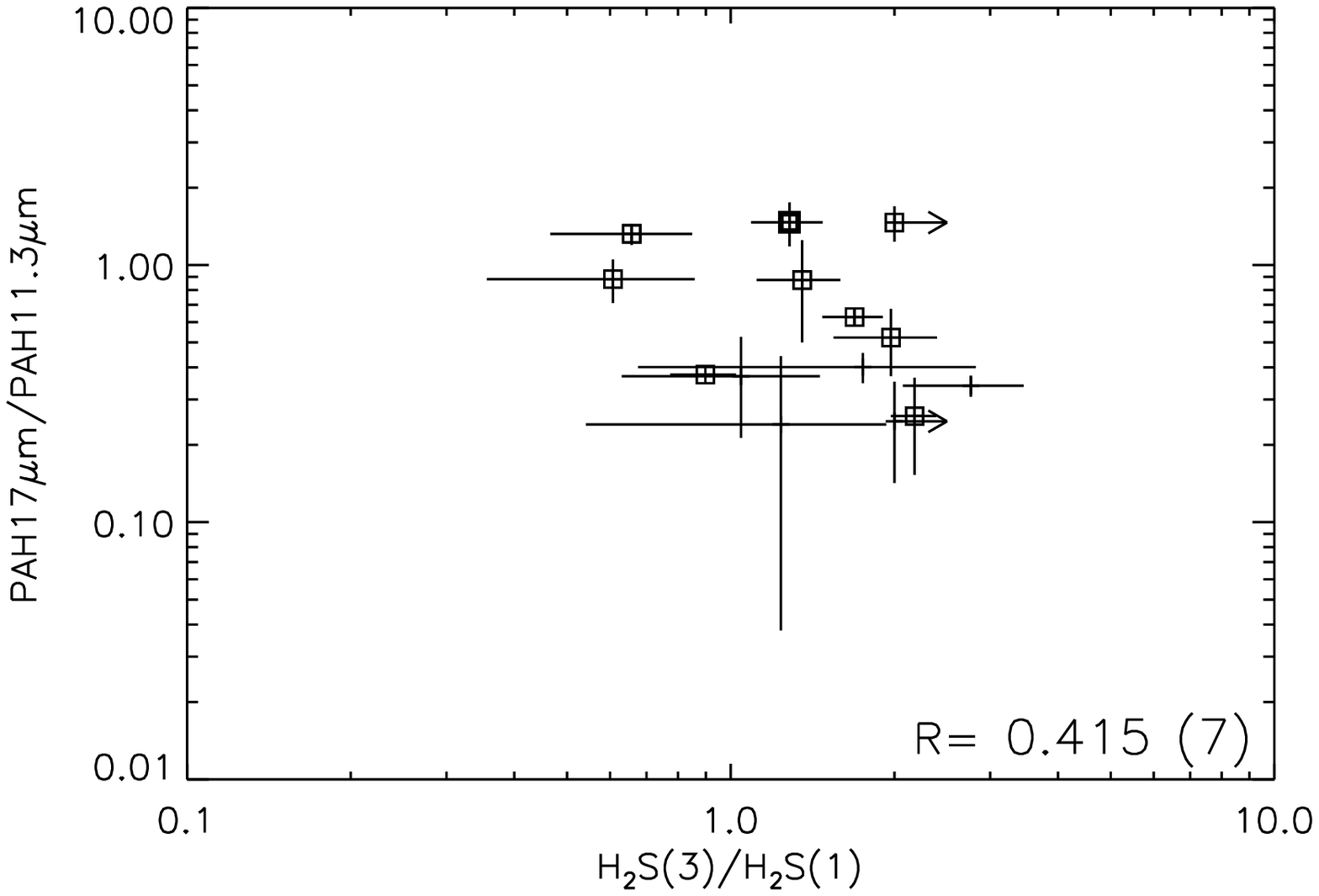}
\caption{Variations of the PAH 7.7$^*$/11.3 and PAH 17/11.3 ratios plotted against the line ratios of [NeIII]/[NeII] and H$_2$S(3)/H$_2$S(1). }
\end{figure}    
\clearpage

\begin{figure}
\epsscale{.45}
\plotone{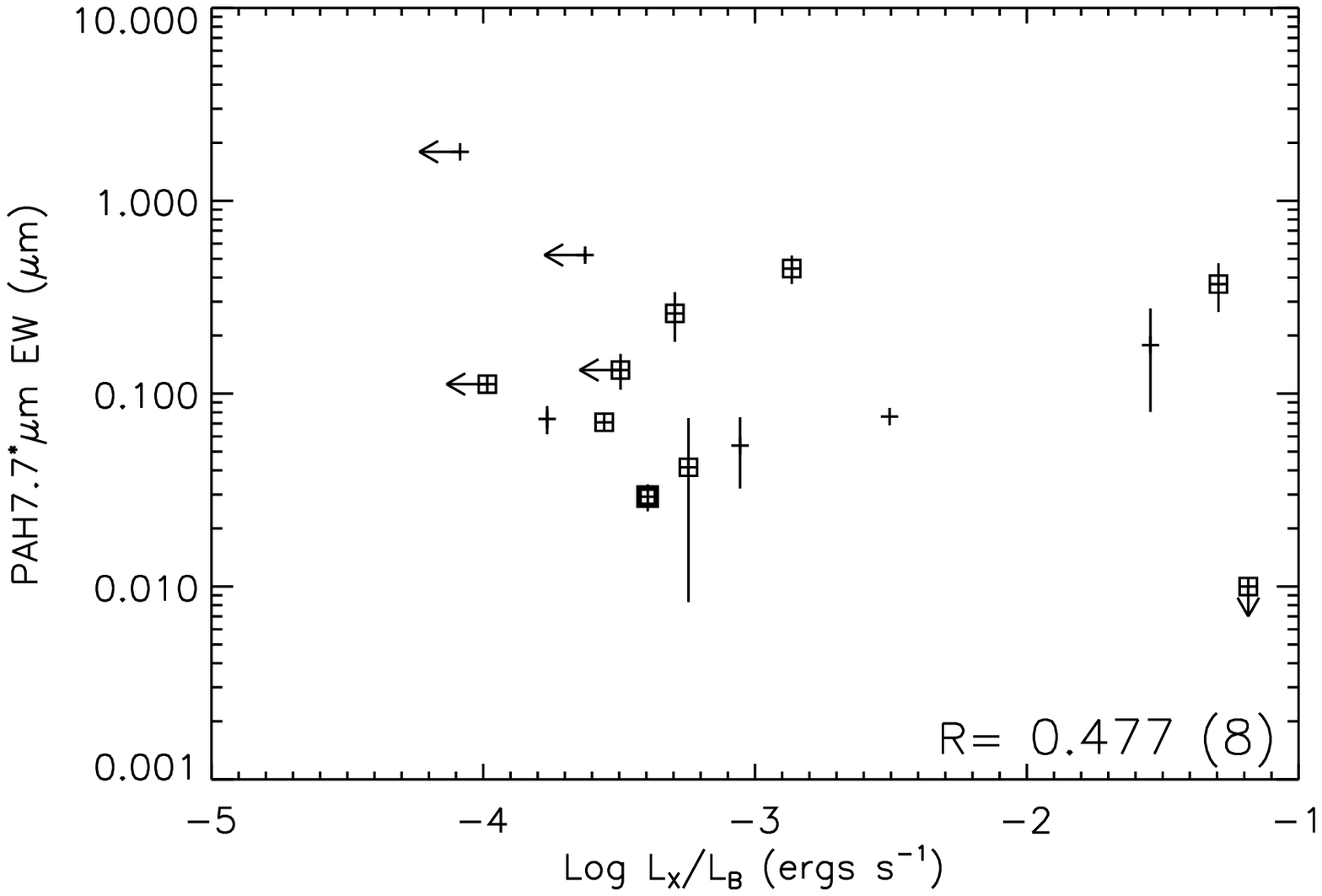}\plotone{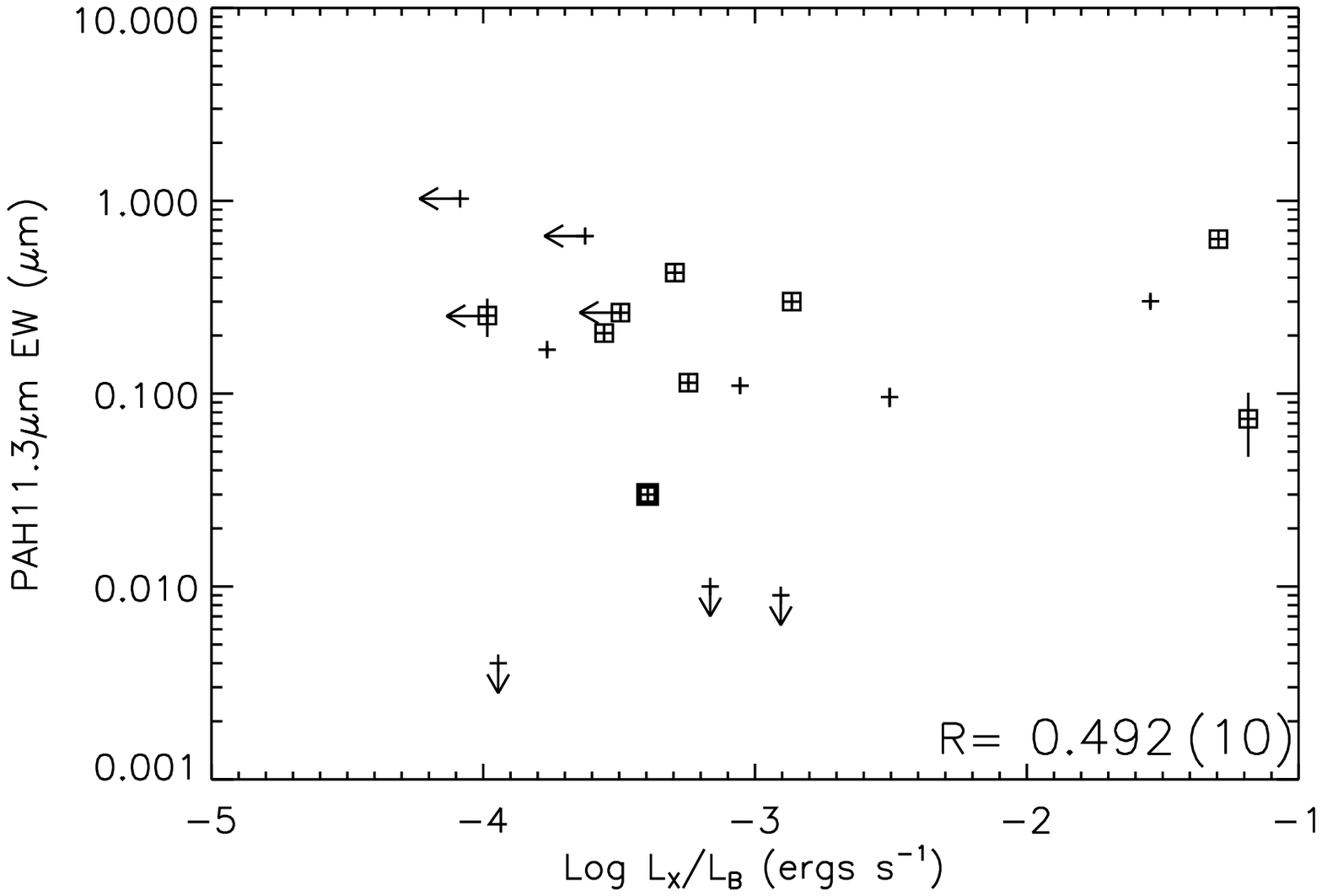}\\
\plotone{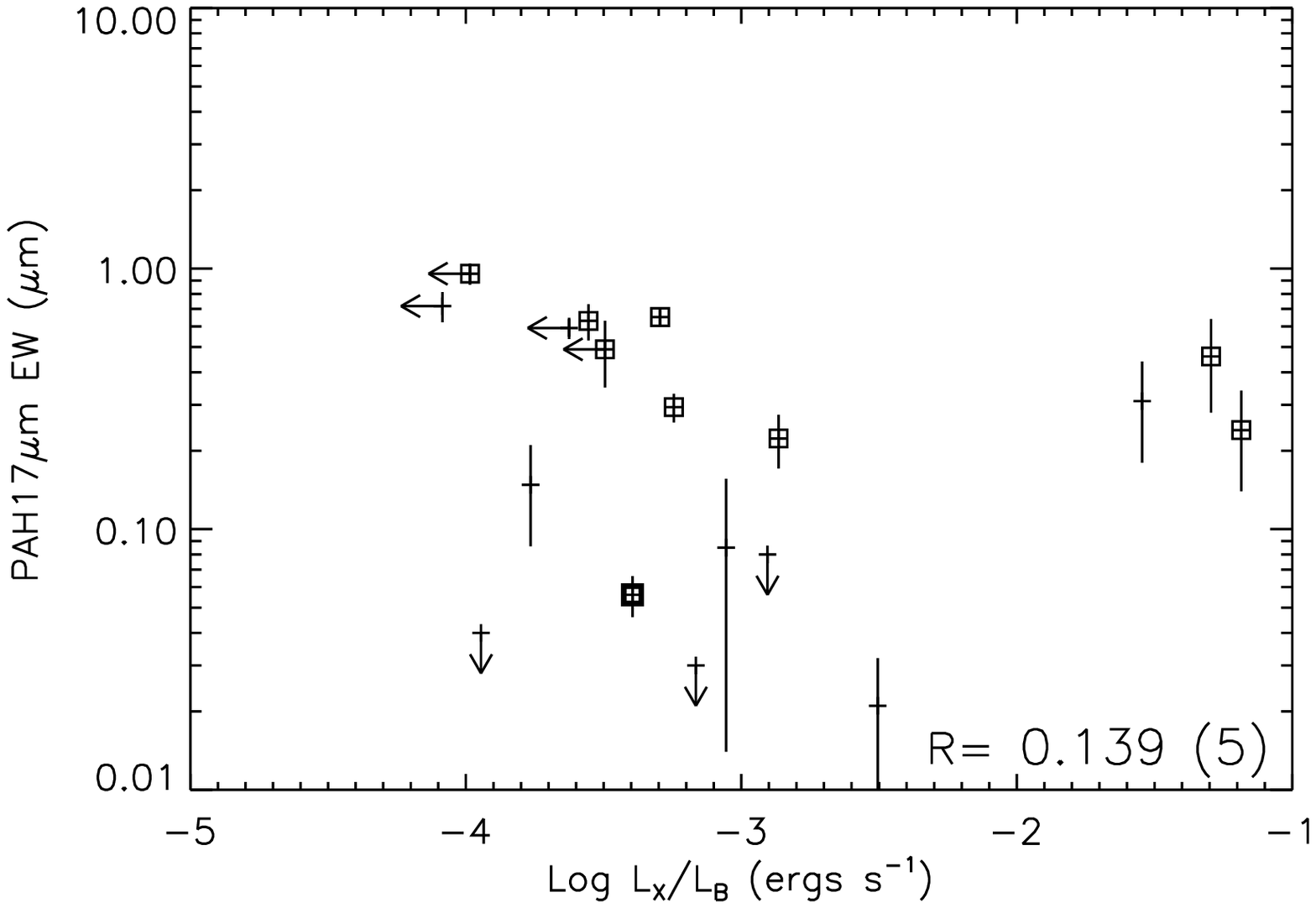}\plotone{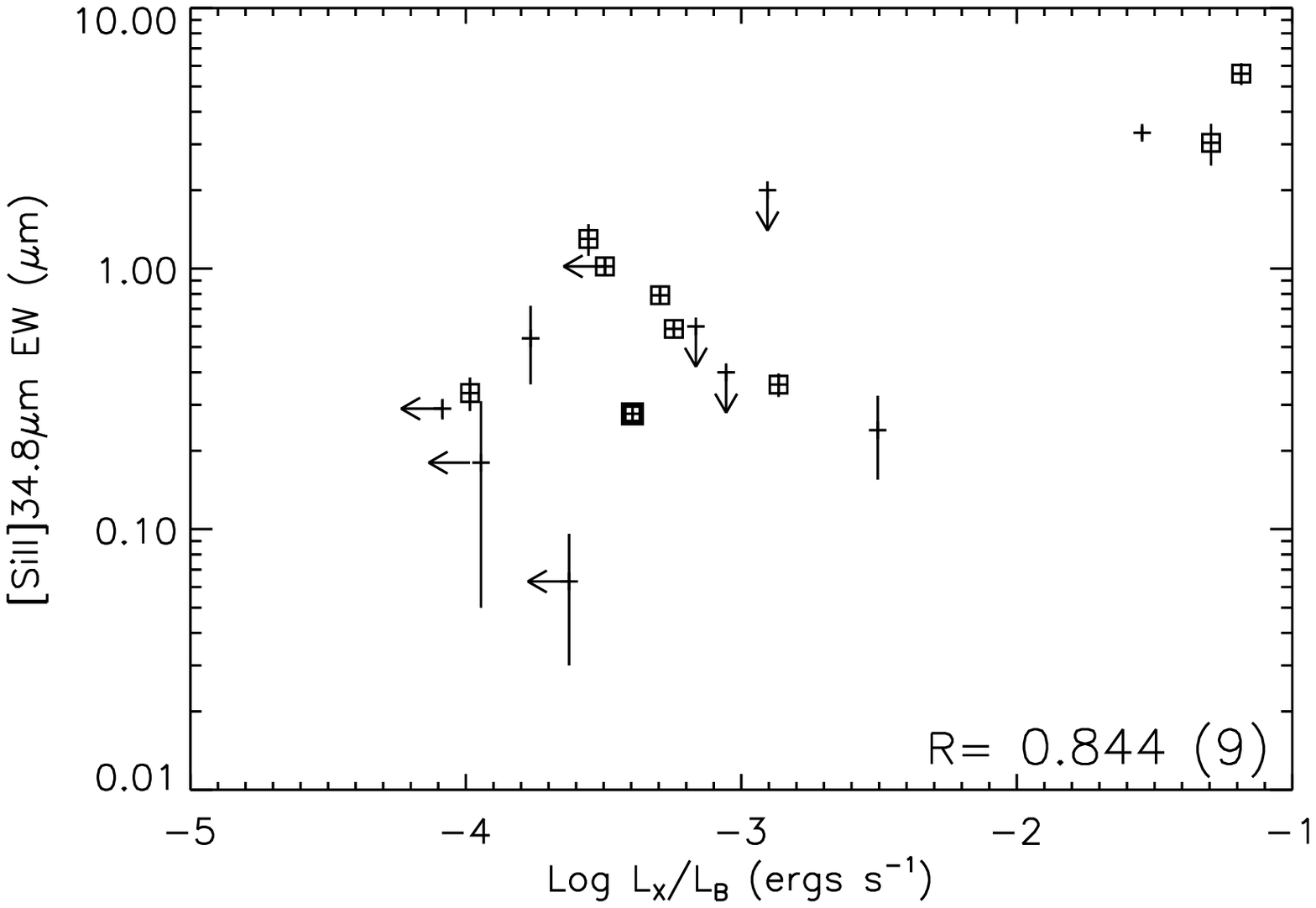}
\caption{Equivalent widths of the PAH 7.7$^*$ $\mu$m, 11.3 $\mu$m, and 17 $\mu$m features as well as that of the [SiII] line plotted as a function of the X-ray relative to the blue luminosity.}
\end{figure}




\end{document}